# Series solutions of Laguerre- and Jacobi-type differential equations in terms of orthogonal polynomials and physical applications

A. D. Alhaidari

*Saudi Center for Theoretical Physics, P. O. Box 32741, Jeddah 21438, Saudi Arabia*

**Abstract**: We introduce two ordinary second-order linear differential equations of the Laguerre- and Jacobi-type. Solutions are written as infinite series of square integrable functions in terms of the Laguerre and Jacobi polynomials, respectively. The expansion coefficients of the series satisfy three-term recursion relations, which are solved in terms of orthogonal polynomials with continuous and/or discrete spectra. Most of these are well-known polynomials whereas few are not. We present physical applications of these differential equations in quantum mechanics.

## 1. Introduction

Differential equations and their solutions might be the most popular mathematical tool used in science and engineering to model systems and processes (see, for example, Refs. [1-3]). Over time, an extensive list of differential equations and their solutions has been created. A differential equation can be ordinary, partial, differential-algebraic, homogeneous or heterogeneous. Typically, differential equations are classified according to linearity, order and degree. An ordinary differential equation (ODE) along with its data (e.g., initial values, boundary conditions, etc.) is said to represent a "well-posed problem" if a unique solution exists with continuous dependence on the data. In this work, we introduce two linear second order ODEs and obtain their solutions as infinite bounded series. These two equations were encountered recently while applying the tridiagonal representation approach (TRA) in the solution of various quantum mechanical problems [4-5], which lead to a formulation of quantum mechanics based on orthogonal polynomials without the need to specify a potential function [6-8]. All properties of the physical system (e.g., bound states energies, scattering phase shift, density of states, etc.) are derived from the properties of these polynomials (e.g., their spectrum formula, weight function, asymptotics, etc.) [7-8]. The first ODE is the following 5-parameter Laguerre-type differential equation:

$$\left[x\frac{d^2}{dx^2}+\left(a+bx\right)\frac{d}{dx}+A_+x+\frac{A_-}{x}\right]y(x)=A_0y(x). \qquad (1)$$

where $\{a,b,A_\pm,A_0\}$ are real parameters. We refer to it as Laguerre-type because the factors multiplying the first and second order derivatives have the same structure as that of the Laguerre differential equation. We propose a solution written as the following infinite sum

$$y(x)=\sum_{n=0}^{\infty}f_n\phi_n(x), \qquad (2)$$

where $\{\phi_n(x)\}$ is a complete set of functions that are square integrable with respect to some proper integration measure, $d\zeta(x)$. That is, we assume $\langle\phi_n|\phi_m\rangle \equiv \int \phi_n(x)\phi_m(x)d\zeta(x) = \delta_{nm}$. Therefore, the series (2) is bounded if the norm

$$\|y\|^2 = \int y(x)^2 d\zeta(x) = \sum_{n=0}^{\infty} f_n^2 , \tag{3}$$

is finite. The differential equation (1) has a regular singularity at $x=0$ and an irregular (essential) singularity at $x \to +\infty$. Thus, due to Fuchs' theorem and using Frobenius method, we can propose the following basis elements:

$$\phi_n(x) = c_n x^{\alpha} e^{-\beta x} L_n^{\nu}(x), \tag{4}$$

where $L_n^{\nu}(x)$ is the Laguerre polynomial of degree *n* in *x* with $x \ge 0$ and $c_n = \sqrt{\frac{\Gamma(n+1)}{\Gamma(n+\nu+1)}}$. The real dimensionless parameters $\{\alpha,\beta,\nu\}$ will be related to the parameters of the differential equation by a certain requirement on the expansion coefficients $\{f_n\}$ that will become clear shortly below. Moreover, the orthogonality measure for this basis is $d\zeta(x) = x^{\nu-2\alpha} e^{(2\beta-1)x} dx$.

We also introduce the following 6-parameter Jacobi-type differential equation:

$$\left\{\left(1-x^2\right)\frac{d^2}{dx^2} - \left[a-b+x\left(a+b\right)\right]\frac{d}{dx} + \frac{A_+}{1+x} + \frac{A_-}{1-x} + A_1 x\right\} y(x) = A_0 y(x). \tag{5}$$

where $\{a,b,A_\pm,A_1,A_0\}$ are real parameters. We refer to this ODE as Jacobi-type because the factors multiplying the first and second order derivatives are of the same form as that of the Jacobi differential equation. It has two regular singularities at $x = \pm 1$. Using Frobenius method, we can propose for this case the following square integrable basis elements to be used in the series solution (2):

$$\phi_n(x) = c_n (1-x)^{\alpha} (1+x)^{\beta} P_n^{(\mu,\nu)}(x), \tag{6}$$

where $P_n^{(\mu,\nu)}(x)$ is the Jacobi polynomial of degree *n* in *x* with $-1 \le x \le +1$ and $c_n = \sqrt{\frac{2n+\mu+\nu+1}{2^{\mu+\nu+1}} \frac{\Gamma(n+1)\Gamma(n+\mu+\nu+1)}{\Gamma(n+\mu+1)\Gamma(n+\nu+1)}}$. The real dimensionless parameters $\{\alpha,\beta,\mu,\nu\}$ will be related to the differential equation parameters by a requirement on the expansion coefficients $\{f_n\}$ that will be outlined below. The orthogonality measure for the basis (6) is $d\zeta(x) = (1-x)^{\mu-2\alpha}$ $(1+x)^{\nu-2\beta} dx$.

The differential properties of the Laguerre and Jacobi polynomials together with their recursion relations make it possible (in principle) for the action of the differential operator in (1) and (5) on the respective basis element $\phi_n(x)$ to produce terms proportional to $\phi_n(x)$ and to $\phi_{n\pm1}(x)$. For that to happen, certain relationships among the parameters of the bases and the differential equations and constraints thereof must be satisfied. If so, then substituting *y*(*x*) given by (2) in the differential equations (1) or (5) will result in a three-term recursion relation for the expansion coefficients $\{f_n\}$. To take full advantage of the many powerful tools and analytic properties of orthogonal polynomials in the solution, we then require that the expansion coefficients satisfy the following symmetric three-term recursion relation ($ST^2R^2$) [9-11]

$$z\, f_n(z) = s_n f_n(z) + t_{n-1} f_{n-1}(z) + t_n f_{n+1}(z)\,, \qquad (7)$$

for $n = 1,2,3,...$ and where *z* is some proper function of the differential equation parameters. The recursion coefficients $\{s_n, t_n\}$ depend on the parameters and on *n* but are independent of *z* and such that $t_n^2 > 0$ for all *n*. Therefore, the solution $f_n(z)$ of (7) becomes a polynomial of degree *n* in *z* modulo an overall factor that depends on *z* but is independent of *n*. That is, we write $f_n(z) = p(z)P_n(z)$, where $p(z) = f_0(z)$ making $P_0(z) = 1$. Defining $t_{-1} = 0$ in (7) for $n = 0$ gives $P_1(z) = (z - s_0)/t_0$. The polynomials $\{P_n(z)\}$ are determined explicitly for any degree by the $ST^2R^2$ (7) starting with the initial values $P_0(z)$ and $P_1(z)$. Consequently, the solution of the second order differential equation becomes equivalent to the solution of the $ST^2R^2$ (7). Moreover, it is worthwhile making the following remark about the boundedness of the solution series (2). Writing $y(x,z) = p(z)\sum_{n=0}^{\infty} P_n(z)\phi_n(x)$ then Eq. (3) dictates that $p^2(z)\sum_{n=0}^{\infty} P_n^2(z)$ should be finite. Now, the diagonal Christoffel-Darboux identity associated with the polynomials $\{P_n(z)\}$ is (see, for example, Theorem 2.2.2 in [11])

$$\sum_{n=0}^{N-1} P_n^2(z) = t_{N-1}\left[P_N'(z)P_{N-1}(z) - P_N(z)P_{N-1}'(z)\right],$$

where *N* is a non-negative integer and the prime stands for the derivative with respect to *z*. Therefore, the infinite limit of this identity along with the asymptotics ($n \to \infty$) of $P_n(z)$ guarantee boundedness of the series provided that $p^2(z)$ is proportional to their positive definite weight (i.e., $\int p^2(z)P_n(z)P_m(z)dz \propto \delta_{nm}$).

In Appendix A and B, we derive sets of $ST^2R^2$s associated with the Laguerre-type and Jacobi-type differential equations, respectively. In sections 2, we show that the polynomials satisfying the recursion relations associated with the solution of Eq. (1) are either the Meixner-Pollaczek polynomials or the continuous dual Hahn polynomials depending on the choice of values of the parameters. However, if one or more of the parameters is discrete (countably infinite or finite) then the solution will be written in terms of the discrete version of these polynomials. That is, in terms of the Meixner or Krawtchouk polynomials for the former case and the dual Hahn polynomial for the latter. On the other hand, in sections 3 we show that if $A_1 = 0$ in Eq. (5) then the polynomial solution of the corresponding $ST^2R^2$ is either the Wilson polynomial or its discrete version the Racah polynomial depending on the nature of the parameters in the equation. However, if $A_1 \neq 0$ then the solution will be written in terms of a new orthogonal polynomial with continuous spectrum or one of its two discrete versions with finite or infinite spectra. Finally, in section 6, we give physical applications for the solution of these two differential equations in quantum mechanics where we write the equivalent Schrödinger equation, identify the corresponding potential functions, and obtain the associated energy spectra for the bound states and/or the phase shift for the scattering states.

# 2. Solutions of the Laguerre-type differential equation

In this section, we obtain the solutions of Eq. (1) by identifying the orthogonal polynomials associated with the $ST^2R^2$s (A8a) and (A8b) subject to the constraints (A7a) and (A7b), respectively. We consider the continuous spectrum case as well as the discrete spectrum. From this point forward and in all subsequent sections, we deal with the normalized version of orthogonal polynomials where the corresponding three-term recursion relation is symmetric and the associated orthogonality relation is of the form $\int_\Omega \rho(z)P_n(z)P_m(z)dz = \delta_{nm}$ for the continuous spectrum or $\sum_k \rho(k)P_n(k)P_m(k) = \delta_{nm}$ for the discrete spectrum, with $\rho(z)$ and $\rho(k)$ being the normalized continuous and normalized discrete weight function, respectively.

## 2.1 Solutions of the $ST^2R^2$ (A8a)

In the following subsections, we identify the orthogonal polynomials in the solution of the first $ST^2R^2$ associated with the Laguerre-type differential equation.

**2.1.1 The two-parameter Meixner-Pollaczek polynomial:**
We start by comparing the $ST^2R^2$ (A8a) to that of the (normalized) Meixner-Pollaczek polynomial $P_n^{\frac{\nu+1}{2}}(z,\theta)$ that reads (see Eq. 9.7.3 in [12])

$$\begin{aligned}(2z\sin\theta)P_n^{\frac{\nu+1}{2}}(z,\theta) &= -\left[(2n+\nu+1)\cos\theta\right]P_n^{\frac{\nu+1}{2}}(z,\theta)\\ &+\sqrt{n(n+\nu)}P_{n-1}^{\frac{\nu+1}{2}}(z,\theta)+\sqrt{(n+1)(n+\nu+1)}P_{n+1}^{\frac{\nu+1}{2}}(z,\theta)\end{aligned} \tag{8}$$

where the normalized version of the polynomial is (see Eq. 9.7.1 in [12])

$$P_n^{\mu}(z,\theta)=\sqrt{\frac{(2\mu)_n}{n!}}\,e^{\mathrm{i}n\theta}\,{}_2F_1\left(\left.{}^{-n,\mu+\mathrm{i}z}_{\;\;2\mu}\right|1-e^{-2i\theta}\right), \tag{9}$$

where $z\in\mathbb{R}$, $\mu>0$, $0\le\theta\le\pi$ and $(a)_n = a(a+1)(a+2)...(a+n-1)=\frac{\Gamma(n+a)}{\Gamma(a)}$. The comparison gives the following polynomial parameters assignments and constraint in addition to (A7a)

$$\cos\theta=\frac{4A_+-b^2-1}{4A_+-b^2+1},\qquad z=-\frac{A_0+\frac{1}{2}ab}{\sqrt{4A_+-b^2}},\qquad 4A_+\ge b^2. \tag{10}$$

Therefore, the solution of the Laguerre-type differential equation (1) is written for a given value of $z$ as

$$y(z,x)=p(z)\sum_{n=0}^{\infty}P_n^{\frac{\nu+1}{2}}(z,\theta)\phi_n(x), \tag{11}$$

where $\phi_n(x)$ is given by (4) and $p(z)$ is a normalization factor, which is required by the boundedness of the series to be proportional to the square root of the positive definite weight function of $P_n^{\frac{\nu+1}{2}}(z,\theta)$ that reads (see Eq. 9.7.2 in [12])

$$p^2(z)=\frac{1/2\pi}{\Gamma(\nu+1)}\left(2\sin\theta\right)^{\nu+1}e^{(2\theta-\pi)z}\left|\Gamma\left(\tfrac{\nu+1}{2}+\mathrm{i}z\right)\right|^2. \tag{12}$$

### 2.1.2 The discrete Meixner polynomial:

For discrete parameter(s) that can assume countably infinite values, we compare the ST$^2$R$^2$ (A8a) to that of the Meixner polynomial $M_n^{\frac{\nu+1}{2}}(k,\tau)$. The ST$^2$R$^2$ of the (normalized) Meixner polynomial reads as follows (see Eq. 9.10.3 in [12])

$$
\begin{aligned}
(\tau-1)k\,M_n^{\frac{\nu+1}{2}}(k,\tau) &= -\left[n(1+\tau)+(\nu+1)\tau\right]M_n^{\frac{\nu+1}{2}}(k,\tau) \\
&+\sqrt{n(n+\nu)\tau}\,M_{n-1}^{\frac{\nu+1}{2}}(k,\tau)+\sqrt{(n+1)(n+\nu+1)\tau}\,M_{n+1}^{\frac{\nu+1}{2}}(k,\tau)
\end{aligned}
\tag{13}
$$

where the normalized version of the polynomial is (see Eq. 9.10.1 in [12])

$$
M_n^{\mu}(k,\tau)=\sqrt{\frac{(2\mu)_n}{n!}}\,\tau^{n/2}\,{}_2F_1\left(\begin{matrix}-n,-k\\2\mu\end{matrix}\middle|1-\tau^{-1}\right), \tag{14}
$$

where $\tau=e^{-2\theta}$ with $\theta>0$, and $k\in\mathbb{N}_0$. Comparing this with the ST$^2$R$^2$ (A8a) gives the following polynomial parameters assignments and constraint in addition to (A7a)

$$
\cosh\theta=\frac{4A_+-b^2-1}{4A_+-b^2+1},\quad 2A_0+ab=\frac{-\sinh\theta}{(\cosh\theta)-1}\left(2k+\nu+1\right),\qquad 4A_+\le b^2-1\,. \tag{15}
$$

Since (A7a) requires $\nu^2=(1-a)^2-4A_-$, then the second equation in (15) above implies that at least one of the two parameters $A_0$ or $A_-$ is discrete via its dependence on the index *k*. So, either $A_0$ is linear in *k* or $A_-$ is quadratic in *k*. The choice depends on the particular case under study. Consequently, the solution of the Laguerre-type differential equation (1) is given for a fixed value of the non-negative integer *k* as

$$
y(k,x)=p(k)\sum_{n=0}^{\infty}M_n^{\frac{\nu+1}{2}}(k,\tau)\phi_n(x)\,, \tag{16}
$$

where $\phi_n(x)$ is as shown in (4) and *p*(*k*) is a normalization factor, which is chosen by the requirement of finiteness of the series as the square root of the following positive definite discrete weight function of $M_n^{\frac{\nu+1}{2}}(k,\tau)$ (see Eq. 9.10.2 in [12])

$$
p^2(k)=(1-\tau)^{\nu+1}\,\frac{\Gamma(k+\nu+1)\tau^k}{\Gamma(\nu+1)\Gamma(k+1)}\,. \tag{17}
$$

### 2.1.3 The discrete Krawtchouk polynomial:

If the parameters are discrete with a finite number of values, then we compare (A8a) with the ST$^2$R$^2$ for the (normalized) Krawtchouk polynomial that reads (see Eq. 9.11.3 in [12])

$$
\begin{aligned}
k\,K_n^N(k,\tau) &= \left[N\tau+n(1-2\tau)\right]K_n^N(k,\tau) \\
&-\sqrt{n(N-n+1)\tau(1-\tau)}\,K_{n-1}^N(k,\tau)-\sqrt{(n+1)(N-n)\tau(1-\tau)}\,K_{n+1}^N(k,\tau)
\end{aligned}
\tag{18}
$$

where the normalized version of the polynomial is (see Eq. 9.11.1 in [12])

$$
K_n^N(k,\tau)=\sqrt{\frac{N!}{n!(N-n)!}}\left(\frac{\tau}{1-\tau}\right)^{n/2}{}_2F_1\left(\begin{matrix}-n,-k\\-N\end{matrix}\middle|\tau^{-1}\right), \tag{19}
$$

where $n,k=0,1,..,N$, and $\tau=\frac{1}{2}\left(1+\tanh\theta\right)$ with $\theta\in\mathbb{R}$. We rewrite the first and second square roots in (18) as

$$\sqrt{n(N-n+1)\tau(1-\tau)}=+\mathrm{i}\sqrt{\tau(1-\tau)}\sqrt{n(n-N-1)}\,, \tag{20a}$$

$$\sqrt{(n+1)(N-n)\tau(1-\tau)}=-\mathrm{i}\sqrt{\tau(1-\tau)}\sqrt{(n+1)(n-N)}\,. \tag{20b}$$

where we have used $\sqrt{-1}=\pm\mathrm{i}$. Dividing both sides of (18) by $\sqrt{\tau(1-\tau)}$, we obtain

$$\begin{aligned}\frac{k}{\sqrt{\tau(1-\tau)}}\tilde{K}_n^N\left(k,\tau\right)=\frac{1}{\sqrt{\tau(1-\tau)}}\left[N\tau+n\left(1-2\tau\right)\right]\tilde{K}_n^N\left(k,\tau\right)\\ -\sqrt{n(n-N-1)}\tilde{K}_{n-1}^N\left(k,\tau\right)-\sqrt{(n+1)(n-N)}\tilde{K}_{n+1}^N\left(k,\tau\right)\end{aligned} \tag{21}$$

where $\tilde{K}_n^N\left(k,\tau\right)=\mathrm{i}^n K_n^N\left(k,\tau\right)$. Comparing this with the ST$^2$R$^2$ (A8a), we obtain the following parameter assignments in addition to (A7a)

$$\nu=-N-1, \qquad \sinh\theta=-\frac{4A_+-b^2-1}{4A_+-b^2+1}, \tag{22a}$$

$$2A_0+ab=\frac{1}{1-\sinh\theta}\left[2k\cosh\theta-N\left(\sinh\theta+\sqrt{\frac{1+\tanh\theta}{1-\tanh\theta}}\right)\right]. \tag{22b}$$

Equation (22b) means that $A_0$ is discrete with a linear dependence on *k*. Consequently, the solution of the Laguerre-type differential equation (1) for this case is given for a fixed value of *k* (from within the range $k=0,1,..,N$) as

$$y(k,x)=p(k)\sum_{n=0}^{N}\mathrm{i}^n K_n^N\left(k,\tau\right)\phi_n(x)\,, \tag{23}$$

where $\phi_n(x)$ is as shown in (4) and *p*(*k*) is a normalization factor, which is the square root of the discrete weight function for $K_n^N\left(k,\tau\right)$ that reads (see Eq. 9.11.2 in [12])

$$p^2(k)=(1-\tau)^{N-k}\frac{\Gamma\left(N+1\right)\tau^k}{\Gamma\left(N-k+1\right)\Gamma\left(k+1\right)}\,. \tag{24}$$

Note that the value of the index $\nu$ in $L_n^\nu(x)$ as $\nu=-N-1$ is acceptable since $n\le N$. Moreover, the constraint $\nu^2=(1-a)^2-4A_-$ and $\nu=-N-1$ dictate that this solution is acceptable only if the two differential equation parameters *a* and $A_-$ are chosen such that the value $\sqrt{(1-a)^2-4A_-}$ $-1$ is a non-negative integer, which is equal to the size of the spectrum *N*.

## 2.2 Solutions of the ST$^2$R$^2$ (A8b)

In the following subsections, we consider the second ST$^2$R$^2$ associated with the Laguerre-type differential equation and identify the corresponding orthogonal polynomials.

**2.2.1 The three-parameter dual Hahn polynomial:**

We compare (A8b) to the ST$^2$R$^2$ of the (normalized) continuous dual Hahn polynomial $S_n^\tau\left(z^2;\frac{\nu+1}{2},\frac{\nu+1}{2}\right)$ that reads (see Eq. 9.3.4 in [12])

$$\begin{aligned} z^2 S_n^\tau &= \left[\left(n+\tau+\tfrac{\nu+1}{2}\right)^2 + n(n+\nu) - \tau^2\right] S_n^\tau \\ &-\left(n+\tau+\tfrac{\nu-1}{2}\right)\sqrt{n(n+\nu)}\, S_{n-1}^\tau - \left(n+\tau+\tfrac{\nu+1}{2}\right)\sqrt{(n+1)(n+\nu+1)}\, S_{n+1}^\tau \end{aligned} \tag{25}$$

where the normalized version of the polynomial is (see Eq. 9.3.1 in [12])

$$S_n^\tau(z^2;a,b) = \sqrt{\frac{(\tau+a)_n(\tau+b)_n}{n!\,(a+b)_n}}\; {}_3F_2\left(\left.\begin{matrix}-n,\tau+iz,\tau-iz\\ \tau+a,\tau+b\end{matrix}\right|1\right), \tag{26}$$

where $z>0$ and $\mathrm{Re}(\tau,a,b)>0$ with non-real parameters occurring in conjugate pairs. The comparison gives the following polynomial parameters assignments in addition to (A7b)

$$\tau = A_0 + \frac{ab+1}{2}, \quad z^2 = A_- - \frac{1}{4}(a-1)^2, \qquad 4A_- \ge (a-1)^2. \tag{27}$$

For a pure continuous spectrum, we require $\tau>0$, which means that $A_0 > -\frac{ab+1}{2}$ and the solution for a given *z* is written as

$$y(z,x) = p(z)\sum_{n=0}^{\infty} S_n^\tau\left(z^2;\tfrac{\nu+1}{2},\tfrac{\nu+1}{2}\right)\phi_n(x), \tag{28}$$

where $\phi_n(x)$ is given by Eq. (4) and *p*(*z*) is a normalization factor, which is the square root of the weight function associated with $S_n^\tau\left(z^2;\frac{\nu+1}{2},\frac{\nu+1}{2}\right)$ that reads (see Eq. 9.3.2 in [12])

$$p^2(z) = \frac{1}{2\pi}\frac{\left|\Gamma(\tau+iz)\Gamma\left(\frac{\nu+1}{2}+iz\right)^2/\Gamma(2iz)\right|^2}{\Gamma\left(\tau+\frac{\nu+1}{2}\right)^2\Gamma(\nu+1)}. \tag{29}$$

On the other hand, if $\tau<0$ then the spectrum is a mix of a continuous part and a discrete part of finite size *N*, which is the largest integer less than or equal to $-\tau$. That is, the variable *z* will be the union of a continuous set $z\in\mathbb{R}^+$ and a discrete finite set $z\in\{z_k = k+\tau\}_{k=0}^{N}$. In that case, the solution for a given *k* and *z* is written as the following two sums

$$y(z,k,x) = p(z)\sum_{n=0}^{\infty} S_n^\tau\left(z^2;\tfrac{\nu+1}{2},\tfrac{\nu+1}{2}\right)\phi_n(x) + p(k)\sum_{n=0}^{N} S_n^\tau\left(-z_k^2;\tfrac{\nu+1}{2},\tfrac{\nu+1}{2}\right)\phi_n(x), \tag{30}$$

where $p(k) = -2\dfrac{\Gamma\left(\frac{\nu+1}{2}-\tau\right)^2}{\Gamma(\nu+1)\Gamma(1-2\tau)}\left[(-1)^k(k+\tau)\left(\tfrac{\nu+1}{2}+\tau\right)_k^2(2\tau)_k\Big/\left(\tfrac{1-\nu}{2}+\tau\right)_k^2 k!\right]$ (see Eq. 9.3.3 in [12]).

**2.2.2 The discrete dual Hahn polynomial:**
For a pure discrete spectrum of finite size, we compare the ST²R² (A8b) with that of the (normalized) dual Hahn polynomial $R_n^N\left(z_k^2;\tau,\sigma\right)$ that reads (see Eq. 9.6.3 in [12])

$$\begin{aligned}\left(k+\tfrac{\tau+\sigma+1}{2}\right)^2 R_n^N &= \left[(n+\tau+1)(N-n)+n(N+\sigma+1-n)+\tfrac{1}{4}(\tau+\sigma+1)^2\right]R_n^N \\ &+\sqrt{n(n+\tau)(N-n+1)(N-n+\sigma+1)}R_{n-1}^N+\sqrt{(n+1)(n+\tau+1)(N-n)(N-n+\sigma)}R_{n+1}^N\end{aligned} \tag{31}$$

where the normalized version of the polynomial is (see Eq. 9.6.1 in [12])

$$R_n^N\left(z_k^2;\tau,\sigma\right)=\sqrt{\frac{(\tau+1)_n(N-n+1)_n}{n!\,(N+\sigma-n+1)_n}}\;{}_3F_2\left(\left.\begin{matrix}-n,-k,k+\tau+\sigma+1\\ \tau+1,-N\end{matrix}\right|1\right), \tag{32}$$

where $n,k=0,1,2,..,N$, $z_k=k+\frac{\tau+\sigma+1}{2}$ and either $\tau,\sigma>-1$ or $\tau,\sigma<-N$. We rewrite the terms inside the square roots by the replacement $-n\to n$ and thus obtain by comparison the following parameter assignments in addition to (A7b)

$$2\tau=\left(2A_0+ab\right)-N-1,\qquad 2\sigma=-\left(2A_0+ab\right)-N-1, \tag{33a}$$

$$\nu=-N-1,\qquad A_-=\tfrac{1}{4}(a-1)^2-\left(k-\tfrac{N}{2}\right)^2. \tag{33b}$$

Therefore, the differential equation parameter $A_-$ is required to be discrete with *k* as shown in (33b). Moreover, the requirement that $\tau,\sigma>-1$ or $\tau,\sigma<-N$ imposes one of two conditions $\pm\left(2A_0+ab\right)>N-1$. Finally, the solution of the Laguerre-type differential equation (1) for this case is written for a fixed value of *k* (from within the range $k=0,1,..,N$) as

$$y(k,x)=p(k)\sum_{n=0}^{N}R_n^N\left(z_k^2;\tau,\sigma\right)\phi_n(x), \tag{34}$$

where $z_k=k-\frac{N}{2}$ and $\phi_n(x)$ is as shown in (4). Here also, the value of the index $\nu$ in $L_n^\nu(x)$ as $\nu=-N-1$ is acceptable since $n\le N$. The normalization factor *p*(*k*) is proportional to the square root of the discrete weight function for $R_n^N\left(z_k^2;\tau,\sigma\right)$ that reads (see Eq. 9.6.2 in [12])

$$p^2(k)=(\sigma+1)_N\frac{(2k+\tau+\sigma+1)(\tau+1)_k(N-k+1)_k}{(k+\tau+\sigma+1)_{N+1}(\sigma+1)_k\,k!}. \tag{35}$$

# 3. Solutions of the Jacobi-type differential equation

In this section, we obtain solutions of Eq. (5) by identifying the orthogonal polynomials associated with the ST²R² (B13a) and (B13c) subject to the constraints (B12a) and (B12c), respectively. We ignore the ST²R² (B13b) and its associated constraints (B12b) since they are easily obtained from the respective relations (B13c) and (B12c) by the parameter map (B14). We consider the continuous spectrum case as well as the discrete spectrum. We will find out that the ST²R² (B13c) corresponds to known orthogonal polynomials of the hypergeometric type. However, the ST²R² (B13a) is associated with a new class of orthogonal polynomials whose properties (weight function, generating function, orthogonality, zeros, asymptotics, Rodrigues-type formula, differential or shift formula, etc.) are yet to be derived in the proper mathematics literature.

## 3.1 Solutions of the ST²R² (B13a)

In the following subsections, we consider the first ST²R² associated with the Jacobi-type differential equation. We start by observing that the asymptotic limit of the recursion coefficients ratio $\lim_{n\to\infty}\left(s_n/t_n\right)$ goes like $n^2$. This is different from all well-known orthogonal polynomials in the mathematics literature. However, in the physics literature and since 2005 a polynomial class with this property has been encountered frequently while solving various quantum mechanical problems [13-19]. Unfortunately, the analytic properties of this class of orthogonal polynomials (such as the weight function, generating function, orthogonality, asymptotics, etc.) are yet to be derived. Up to now, these polynomials are defined only by their ST²R²s, which determine all of them explicitly (albeit not in closed form) to any desired degree starting with their initial values. Nevertheless, due to the lack of knowledge of critical properties of these polynomials, we will not pursue the details of the associated solutions.

**3.1.1 The four-parameter *new* polynomial:**
We compare (B13a) to the ST²R² of the normalized version of the *new* polynomial with continuous spectrum, $H_n^{(\mu,\nu)}(z^{-1};\theta,\sigma)$, which is defined in [4,20] that reads

$$\begin{aligned}\left(\cos\theta\right)H_n^{(\mu,\nu)}(z^{-1};\theta,\sigma)=&\left\{z^{-1}\sin\theta\left[\sigma+\left(n+\tfrac{\mu+\nu+1}{2}\right)^2\right]+C_n\right\}H_n^{(\mu,\nu)}(z^{-1};\theta,\sigma)\\&+D_{n-1}H_{n-1}^{(\mu,\nu)}(z^{-1};\theta,\sigma)+D_nH_{n+1}^{(\mu,\nu)}(z^{-1};\theta,\sigma)\end{aligned}\qquad(36)$$

where $z\in\mathbb{R}$, $0<\theta<\pi$, $C_n$ and $D_n$ are defined in Appendix B by Eq. (B10). Thus, $H_n^{(\mu,\nu)}(z^{-1};\theta,\sigma)$ is a polynomial of degree *n* in $z^{-1}$ and in $\sigma$. Taking $z\to\infty$ turns (36) into the ST²R² of the (normalized) Jacobi polynomial $P_n^{(\mu,\nu)}(\cos\theta)$. The comparison with (B13a) gives the following polynomial parameters assignments in addition to (B12a)

$$\sigma=0,\ \cos\theta=\left[A_0-\tfrac{1}{4}(a+b-1)^2\right]/A_1,\ z=-sign(A_1)\sqrt{A_1^2-\left[A_0-\tfrac{1}{4}(a+b-1)^2\right]^2},\qquad(37)$$

with the parameter constraint $A_1^2\ge\left[A_0-\tfrac{1}{4}(a+b-1)^2\right]^2$. Therefore, the solution of the Jacobi-type differential equation (5) is written for a given value of *z* as

$$y(z,x)=p(z)\sum_{n=0}^{\infty}H_n^{(\mu,\nu)}(z^{-1};\theta,0)\phi_n(x),\qquad(38)$$

where $\phi_n(x)$ is given by (6) and *p*(*z*) is a normalization factor, which is proportional to the square root of weight function of $H_n^{(\mu,\nu)}(z^{-1};\theta,\sigma)$ that is yet to be derived.

**3.1.2 The discrete *new* polynomials:**
If, on the other hand, $A_1^2<\left[A_0-\tfrac{1}{4}(a+b-1)^2\right]^2$ then the spectrum is discrete and we compare the ST²R² (B13a) with that of the discrete version of $H_n^{(\mu,\nu)}(z^{-1};\theta,\sigma)$. There are two discrete versions of this orthogonal polynomial; one with a countably infinite spectrum and another with a finite spectrum. If we refer to both of these collectively as $G_n^{(\mu,\nu)}(k;\tau,\sigma)$, then the corresponding ST[2]R[2] reads as follows [4,20]

$$\begin{aligned}(1+\tau)G_n^{(\mu,\nu)}(k;\tau,\sigma) &= \left\{z_k^{-1}(1-\tau)\left[\sigma+\left(n+\tfrac{\mu+\nu+1}{2}\right)^2\right]+2\sqrt{\tau}C_n\right\}G_n^{(\mu,\nu)}(k;\tau,\sigma)\\ &+2\sqrt{\tau}\left[D_{n-1}G_{n-1}^{(\mu,\nu)}(k;\tau,\sigma)+D_nG_{n+1}^{(\mu,\nu)}(k;\tau,\sigma)\right]\end{aligned} \tag{39}$$

where $k$ is a non-negative integer of either finite or infinite range and $1>\tau>0$. The discrete polynomial parameter $z_k$ is determined from the condition that forces the asymptotics ($n\to\infty$) of $H_n^{(\mu,\nu)}(z^{-1};\theta,\sigma)$ to vanish (i.e., its spectrum formula). However, this asymptotics is not found in the published literature and it is yet to be derived analytically. Had this been known, we would have been able to write the solution of the Jacobi-type differential equation (5) for a fixed value of the non-negative integer $k$ as

$$y(k,x)=p(k)\sum_n G_n^{(\mu,\nu)}(k;\tau,0)\phi_n(x), \tag{40}$$

where $\phi_n(x)$ is given by (6) and $p(k)$ is a normalization factor. For a finite spectrum of size $N$, one or both of the parameters $\mu$ and $\nu$ will be linearly dependent on $N$. The sum in (40) is either infinite or finite of order $N$.

## 3.2 Solutions of the ST[2]R[2] (B13c)

In the following subsections, we consider the last ST[2]R[2] associated with the Jacobi-type differential equation and identify the corresponding orthogonal polynomials in the series solutions.

### 3.2.1 The four-parameter Wilson polynomial:

We start by comparing (B13c) to the ST[2]R[2] of the (normalized) Wilson polynomial $W_n(z^2;a,b,c,d)$ that reads (see Eq. 9.1.4 in [12])

$$\begin{aligned}z^2W_n &= \left[\frac{(n+a+b)(n+a+c)(n+a+d)(n+a+b+c+d-1)}{(2n+a+b+c+d)(2n+a+b+c+d-1)}+\frac{n(n+b+c-1)(n+b+d-1)(n+c+d-1)}{(2n+a+b+c+d-1)(2n+a+b+c+d-2)}-a^2\right]W_n\\ &-\frac{1}{2n+a+b+c+d-2}\sqrt{\frac{n(n+a+b-1)(n+c+d-1)(n+a+c-1)(n+a+d-1)(n+b+c-1)(n+b+d-1)(n+a+b+c+d-2)}{(2n+a+b+c+d-3)(2n+a+b+c+d-1)}}\,W_{n-1}\\ &-\frac{1}{2n+a+b+c+d}\sqrt{\frac{(n+1)(n+a+b)(n+c+d)(n+a+c)(n+a+d)(n+b+c)(n+b+d)(n+a+b+c+d-1)}{(2n+a+b+c+d-1)(2n+a+b+c+d+1)}}\,W_{n+1}\end{aligned} \tag{41}$$

where $z>0$ and $\mathrm{Re}(a,b,c,d)>0$ with non-real parameters occurring in conjugate pairs. Additionally, the normalized version of the Wilson polynomial is written as (see Eq. 9.1.1 in [12])

$$\begin{aligned}&W_n(z^2;a,b,c,d)=\\ &\sqrt{\left(\frac{2n+a+b+c+d-1}{n+a+b+c+d-1}\right)\frac{(a+b)_n(a+c)_n(a+d)_n(a+b+c+d)_n}{(b+c)_n(b+d)_n(c+d)_n n!}}\;{}_4F_3\left(\left.\begin{matrix}-n,n+a+b+c+d-1,a+ix,a-ix\\ a+b,a+c,a+d\end{matrix}\right|1\right)\end{aligned} \tag{42}$$

where ${}_4F_3\left(\begin{matrix}a,b,c,d\\ e,f,g\end{matrix}\middle|z\right)=\sum_{n=0}^{\infty}\frac{(a)_n(b)_n(c)_n(d)_n}{(e)_n(f)_n(g)_n}\frac{z^n}{n!}$. It is a polynomial of degree $n$ in $z^2$. If we reparametrize the polynomial as $W_n(z^2;\sigma+\mathrm{i}\tau,\sigma-\mathrm{i}\tau,\gamma,\gamma)$, then comparing the resulting ST$^2$R$^2$ (41) to (B13a) gives the following parameters assignments and constraint in addition to (B12c)

$$2\sigma=\nu+1,\ 2\gamma=\mu+1,\ 2\tau=\sqrt{4A_0-(a+b-1)^2}\ ,\ 2z^2=A_- -\tfrac{1}{2}\left(a-1\right)^2. \tag{43}$$

For a pure continuous spectrum $A_0\geq\frac{1}{4}(a+b-1)^2$ and $A_-\geq\frac{1}{2}\left(a-1\right)^2$ (i.e. $\tau^2>0$ and $z^2>0$). The solution of the Jacobi-type differential equation (5) in this case and for a given value of $z$ is written as follows

$$y(z,x)=p(z)\sum_{n=0}^{\infty}W_n(z^2;\sigma+\mathrm{i}\tau,\sigma-\mathrm{i}\tau,\gamma,\gamma)\phi_n(x), \tag{44}$$

where $\phi_n(x)$ is given by (6) and $p(z)$ is proportional to the square root of the weight function for $W_n(z^2;\sigma+\mathrm{i}\tau,\sigma-\mathrm{i}\tau,\gamma,\gamma)$ that reads (see Eq. 9.1.2 in [12])

$$p^2(z)=\frac{1}{2\pi}\frac{\left|\Gamma\left(\frac{\mu+1}{2}+\mathrm{i}z\right)\right|^4\left|\Gamma\left[\frac{\nu+1}{2}+\mathrm{i}(z+\tau)\right]\Gamma\left[\frac{\nu+1}{2}+\mathrm{i}(z-\tau)\right]/\Gamma(2iz)\right|^2}{\left[\Gamma(\mu+1)\Gamma(\nu+1)/\Gamma(\mu+\nu+2)\right]\left|\Gamma\left(\frac{\mu+\nu}{2}+1+\mathrm{i}\tau\right)\right|^4}. \tag{45}$$

On the other hand, if $\tau^2<0$, which means that $A_0<\frac{1}{4}(a+b-1)^2$, then the spectrum is a mix of a continuous part and a discrete part of finite size $N$, which is the largest integer less than or equal to $\sqrt{-\tau^2}-\sigma$. That is, the variable $z$ will be the union of a continuous set $z>0$ and a finite discrete set $z\in\left\{z_k=k+\sigma-\sqrt{-\tau^2}\right\}_{k=0}^{N}$. In that case, the solution for a given $k$ and $z$ is written as two sums

$$\begin{aligned}y(z,k,x)=p(z)&\sum_{n=0}^{\infty}W_n(z^2;\sigma+\mathrm{i}\tau,\sigma-\mathrm{i}\tau,\gamma,\gamma)\phi_n(x)\\ +p(k)&\sum_{n=0}^{N}W_n(-z_k^2;\sigma+\mathrm{i}\tau,\sigma-\mathrm{i}\tau,\gamma,\gamma)\phi_n(x)\end{aligned}. \tag{46}$$

where $p(k)=-2\frac{\Gamma(2\sigma+2\gamma)\Gamma(-2\mathrm{i}\tau)\left[\Gamma(\gamma-\sigma-\mathrm{i}\tau)\right]^2}{\Gamma(-2\mathrm{i}\tau-2\sigma+1)\Gamma(2\gamma)\left[\Gamma(\gamma+\sigma-\mathrm{i}\tau)\right]^2}\left\{(k+\sigma+\mathrm{i}\tau)\frac{(2\sigma+2\mathrm{i}\tau)_k(2\sigma)_k\left[(\sigma+\mathrm{i}\tau+\gamma)_k\right]^2}{(2\mathrm{i}\tau+1)_k\left[(\sigma+\mathrm{i}\tau-\gamma+1)_k\right]^2k!}\right\}$ (see Eq. 9.1.3 in [12]).

**3.2.2 The discrete Racah polynomial:**

For a pure discrete spectrum, we compare the ST$^2$R$^2$ (B13c) with that of the (normalized) Racah polynomial $R_n^N\left(z_k^2;\gamma;\sigma,\sigma\right)$ that reads (see Eq. 9.2.3 in [12])

$$
\begin{aligned}
\left(k-\tfrac{N}{2}\right)^2 R_n^N = -&\left[\frac{(n-N)(n+\gamma+1)(n+\sigma+1)(n+\gamma+\sigma+1)}{(2n+\gamma+\sigma+1)(2n+\gamma+\sigma+2)}+\frac{n(n+\gamma)(n+\sigma)(n+\gamma+\sigma+N+1)}{(2n+\gamma+\sigma)(2n+\gamma+\sigma+1)}-\frac{N^2}{4}\right]R_n^N \\
&+\frac{(n+\gamma+1)(n+\sigma+1)}{(2n+\gamma+\sigma+2)}\sqrt{\frac{(n+1)(n-N)(n+\gamma+\sigma+1)(n+\gamma+\sigma+N+2)}{(2n+\gamma+\sigma+1)(2n+\gamma+\sigma+3)}}R_{n+1}^N \\
&+\frac{(n+\gamma)(n+\sigma)}{(2n+\gamma+\sigma)}\sqrt{\frac{n(n-N-1)(n+\gamma+\sigma)(n+\gamma+\sigma+N+1)}{(2n+\gamma+\sigma-1)(2n+\gamma+\sigma+1)}}R_{n-1}^N
\end{aligned}
\qquad (47)
$$

where $z_k^2=\left(k-\frac{N}{2}\right)^2$. The normalized version of the Racah polynomial, $R_n^N\left(z_k^2;\alpha;\beta,\gamma\right)$, is written as (see Eq. 9.2.1 in [12])

$$
\begin{aligned}
&R_n^N(z_k^2;\alpha;\beta,\gamma)= \\
&\sqrt{\frac{2n+\alpha+\beta+1}{n+\alpha+\beta+1}\frac{(-N)_n(\alpha+1)_n(\gamma+1)_n(\alpha+\beta+2)_n}{(\beta+1)_n(\alpha+\beta-\gamma+1)_n(\alpha+\beta+N+2)_n n!}}\times {}_4F_3\left(\left.\begin{matrix}-n,-k,n+\alpha+\beta+1,k-\beta+\gamma-N\\ \alpha+1,\gamma+1,-N\end{matrix}\right|1\right)
\end{aligned}
\qquad (48)
$$

where $z_k^2=\frac{1}{4}\left(N+\beta-\gamma-2k\right)^2$ and $n,k=0,1,2,..,N$. The comparison of (47) to (B13c) gives the following polynomial parameters assignments and constraint in addition to (B12c)

$$
2\sigma=\mu+\nu+\sqrt{(a+b-1)^2-4A_0}\,,\quad 2\gamma=\mu+\nu-\sqrt{(a+b-1)^2-4A_0}\,, \qquad (49a)
$$

$$
N=-\mu-1\,,\qquad A_-=\tfrac{1}{2}\left(a-1\right)^2-\left(k-\tfrac{N}{2}\right)^2. \qquad (49b)
$$

Therefore, it is required that $A_0<\frac{1}{4}(a+b-1)^2$ and $A_-$ is discretized as shown in (49b). Additionally, the solution of the Jacobi-type differential equation (5) in this case is written for a fixed value of *k* (from within the range $k=0,1,..,N$) as

$$
y(k,x)=p(k)\sum_{n=0}^{N}R_n^N\left(z_k^2;\gamma;\sigma,\sigma\right)\phi_n(x)\,, \qquad (50)
$$

where $z_k^2=\frac{1}{4}\left(N-2k\right)^2$ and $\phi_n(x)$ is as shown in (6). Note that the value of the index $\mu$ in $P_n^{(\mu,\nu)}(x)$ as $\mu=-N-1$ is acceptable since $n\le N$. The normalization factor *p*(*k*) is proportional to the square root of the discrete weight function of $R_n^N\left(z_k^2;\gamma;\sigma,\sigma\right)$ as follows (see Eq. 9.2.2 in [12])

$$
p^2(k)=\frac{(-\sigma-N)_N(-\gamma-N)_N}{(-\gamma-\sigma-N-1)_N(-N+1)_N}\left[\frac{2k-N}{k-N}\frac{(-N)_k(\gamma+1)_k(\sigma+1)_k(-N+1)_k}{(-\sigma-N)_k(-\gamma-N)_k(k!)^2}\right]. \qquad (51)
$$

In the process of comparing the ST$^2$R$^2$ (B13c) with either (41) or (47) we employed the following identity, which is valid for all *n* and real parameters $\{\mu,\nu,\chi\}$.

$$
\begin{aligned}
&\frac{(n+\mu+1)(n+\mu+\nu+1)\left[(2n+\mu+\nu+2)^2+\chi\right]}{(2n+\mu+\nu+1)(2n+\mu+\nu+2)}+\frac{n(n+\nu)\left[(2n+\mu+\nu)^2+\chi\right]}{(2n+\mu+\nu)(2n+\mu+\nu+1)}= \\
&-\frac{4n(n+\mu)}{2n+\mu+\nu}+\frac{1}{2}\left[1+\frac{\mu^2-\nu^2}{(2n+\mu+\nu)(2n+\mu+\nu+2)}\right]\left[(2n+\mu+\nu+2)^2+\chi\right]
\end{aligned}
\qquad (52)
$$

## 4. Physical applications

In the atomic units $\hbar = M = 1$, the time-independent Schrödinger equation in one dimension for a potential function $V(r)$ and energy $E$ reads as follows

$$\left[-\frac{1}{2}\frac{d^2}{dr^2}+V(r)-E\right]\psi(r)=0\,, \tag{53}$$

where the configuration space coordinate $r$ is either the whole real line ($-\infty < r < +\infty$), half of the line ($r \geq 0$), or a finite segment thereof ($r_- \leq r \leq r_+$). Note that if we write $V(r)=\tilde{V}(r)+\ell(\ell+1)/2r^2$, then Eq. (53) becomes the Schrödinger equation in three dimensions with spherical symmetry for the radial potential $\tilde{V}(r)$ and angular momentum quantum number $\ell$. Making the coordinate transformation $r \to x(r)$ and defining $\xi(x)=\lambda^{-1}(dx/dr)$, where $\lambda$ is a real positive parameter of inverse length dimension, then the Schrödinger equation (53) in the new dimensionless coordinate $x$ becomes

$$\lambda^2\xi^2\left[\frac{d^2}{dx^2}+\frac{\xi'}{\xi}\frac{d}{dx}-\frac{2}{\lambda^2\xi^2}W(x)\right]\psi=0\,, \tag{54}$$

where $W(x)=V(r)-E$ and the prime stands for the derivative with respect to $x$.

In the following two subsections, we consider two types of coordinate transformation $x(r)$ and identify the potential functions that turn the Schrödinger equation (54) into the Laguerre-type or Jacobi-type differential equations. Subsequently, we use the results in section 3 above to write down the corresponding wavefunctions as the series expansion (2). If the polynomial in the wavefunction expansion is discrete, then we obtain the energy spectrum formula using the association of the discrete polynomial parameters with the energy. On the other hand, for the continuum energy scattering states, the polynomial in the wavefunction expansion is continuous and the scattering phase shift is obtained from the sinusoidal asymptotics ($n \to \infty$) formula of these polynomials [21-23].

### 4.1 The Laguerre-type differential equation

For the Laguerre-type differential equation (1), we require that the coordinate transformation be such that $x(r) \geq 0$ and $x(\xi'/\xi)=a+bx$ whose solution is $\xi(x)=x^a e^{bx}$, where $a$ and $b$ are dimensionless real parameters. Consequently, Eq. (54) becomes

$$\frac{\lambda^2\xi^2}{x}\left[x\frac{d^2}{dx^2}+(a+bx)\frac{d}{dx}-\frac{2}{\lambda^2}x^{1-2a}e^{-2bx}W(x)\right]\psi=0\,. \tag{55}$$

Resulting in the following function

$$W(x)=-\frac{\lambda^2}{2}x^{2a}e^{2bx}\left(A_+-A_0x^{-1}+A_-x^{-2}\right), \tag{56}$$

and turning Schrödinger equation into the differential equation (1) with $y(x)=\psi(r)$. Therefore, the Laguerre-type differential equation (1) is equivalent to the Schrödinger equation in quantum mechanics for the interaction potential *V*(*r*), which is equal to the function (56) modulo an overall additive constant ($-E$). In the following three subsections, we give examples in this class corresponding to $(a,b)=(0,0)$, $(a,b)=\left(\frac{1}{2},0\right)$, and $(a,b)=(1,0)$.

**4.1.1 The Coulomb problem $(a,b)=(0,0)$:**

These values of $a$ and $b$ give $\xi(x)=1$, which means that $x=\lambda r$ and $r\geq 0$. Moreover, the potential function (56) becomes

$$W(x)=-\frac{\lambda^2}{2}\left[A_+-\frac{A_0/\lambda}{r}+\frac{A_-/\lambda^2}{r^2}\right]. \tag{57}$$

This corresponds to the Coulomb problem in three dimensions with an effective potential that includes the orbital term, which reads $V_{\text{eff}}(r)=\frac{\ell(\ell+1)}{2r^2}+\frac{Z}{r}$. Thus, we make the following choice of parameters

$$A_0=2Z/\lambda\,,\qquad A_-=-\ell(\ell+1),\quad A_+=2E/\lambda^2\,, \tag{58}$$

where *Z* is the electric charge. If we identify this problem with the solution in subsection 2.1.1 in terms of the Meixner-Pollaczek polynomial, then using (10) and (A7a) we obtain

$$\nu=\pm(2\ell+1),\ \alpha=\begin{cases}\ell+1\\-\ell\end{cases},\ \beta=\frac{1}{2},\ \cos\theta=\frac{4\kappa^2-\lambda^2}{4\kappa^2+\lambda^2},\ z=-Z/\kappa\,, \tag{59}$$

where $\kappa=\sqrt{2E}$ is the linear momentum wave number. The bottom choice of sign that corresponds to $\nu=-(2\ell+1)$ and $\alpha=-\ell$ is not physically acceptable because then the wave function blows up at the origin. Moreover, the continuum scattering wavefunction (11) becomes

$$\psi(r,E)=p(E)(\lambda r)^{\ell+1}e^{-\lambda r/2}\sum\nolimits_{n=0}^{\infty}c_nP_n^{\ell+1}(z,\theta)L_n^{2\ell+1}(\lambda r)\,, \tag{60}$$

where *p*(*E*) is given by Eq. (12). The asymptotics ($n\to\infty$) of $P_n^{\mu}(z,\theta)$ gives the phase shift $\delta(z)=\arg\Gamma(\mu+\text{i}z)$ (see, for example, Eq. A4 in Appendix A of [7]). Substituting the physical parameters, we obtain

$$\delta(E)=\arg\Gamma(\ell+1-\text{i}\,Z/\kappa)\,, \tag{61}$$

which is exactly that of the Coulomb problem. On the other hand, if we identify this problem with the solution in subsection 2.1.2 in terms of the Meixner polynomial, then using (15) and (A7a) we obtain

$$\nu=\pm(2\ell+1),\ \alpha=\begin{cases}\ell+1\\-\ell\end{cases},\ \beta=\frac{1}{2},\ \cosh\theta=\frac{4\kappa^2+\lambda^2}{4\kappa^2-\lambda^2},\ E_m=-\frac{1}{2}\frac{Z^2}{(m+\ell+1)}, \tag{62}$$

where $\kappa=\sqrt{-2E}$ and $m=0,1,2,\ldots$. The last formula above is exactly that of the energy spectrum for Coulomb problem. The constraint $4A_+\le b^2-1$ in (15) dictates that for a given quantum numbers $\ell$ and $m$, the scale parameter $\lambda$ must be chosen such that $\lambda\le 2|Z|/(m+\ell+1)$. Moreover, the $m^{\text{th}}$ bound state wavefunction (16) reads as follow

$$\psi_m(r)=p(m)\left(\lambda r\right)^{\ell+1}e^{-\lambda r/2}\sum_{n=0}^{\infty}c_n M_n^{\ell+1}(m,\tau)L_n^{2\ell+1}(\lambda r), \tag{63}$$

where $\tau=\left(\frac{2\kappa-\lambda}{2\kappa+\lambda}\right)^2$ and $p(m)$ given by Eq. (17).

### 4.1.2 The isotropic oscillator $(a,b)=\left(\frac{1}{2},0\right)$:

These values of $a$ and $b$ give $\xi(x)=\sqrt{x}$, which gives $x=\left(\lambda r/2\right)^2$ and $r\ge 0$. Moreover, the potential function (56) becomes

$$W(x)=-\frac{\lambda^2}{2}\left[\frac{A_+\lambda^2}{4}r^2-A_0+\frac{4A_-/\lambda^2}{r^2}\right]. \tag{64}$$

This corresponds to the three-dimensional isotropic oscillator with an effective potential that includes the orbital term as $V_{\text{eff}}(r)=\frac{\ell\left(\ell+1\right)}{2r^2}+\frac{1}{2}\omega^2r^2$. Hence, we make the following choice of polynomial parameters

$$A_+=-4\omega^2/\lambda^4\,,\quad A_-=-\tfrac{1}{4}\ell\left(\ell+1\right),\, A_0=-2E/\lambda^2\,, \tag{65}$$

where $\omega$ is the oscillator frequency. If we identify this problem with the solution in subsection 2.1.2 in terms of the Meixner polynomial, then using (15) and (A7a) we obtain

$$\nu=\pm\left(\ell+\tfrac{1}{2}\right),\ \alpha=\begin{cases}\ell+1\\ -\ell\end{cases},\ \beta=\frac{1}{2},\ \cosh\theta=\frac{\left(4\omega/\lambda^2\right)^2+1}{\left(4\omega/\lambda^2\right)^2-1},\ E_m=\omega\left(2m+\ell+\tfrac{3}{2}\right), \tag{66}$$

with $\lambda^2\le 4\omega$. Physical requirements dictates that we dismiss the choice $\nu=-\left(\ell+\frac{1}{2}\right)$ and $\alpha=-\ell$. The last formula above is identical to the energy spectrum of the isotropic oscillator with frequency $\omega$ and angular momentum quantum number $\ell$. Moreover, the $m^{\text{th}}$ bound state wavefunction (16) reads as follow

$$\psi_m(r)=p(m)\left(\lambda r\right)^{\ell+1}e^{-\lambda^2r^2/2}\sum_{n=0}^{\infty}c_n M_n^{\frac{\ell}{2}+\frac{3}{4}}(m,\tau)L_n^{\ell+\frac{1}{2}}(\lambda^2r^2), \tag{67}$$

where $p(m)$ given by Eq. (17) and we have replaced $\lambda$ by $2\lambda$ making $\tau=\left(\frac{\kappa-\lambda}{\kappa+\lambda}\right)^2$ and requiring that $\lambda^2\le\omega$.

### 4.1.3 The one-dimensional Morse oscillator $(a,b)=(1,0)$:

These values of $a$ and $b$ means that $\xi(x)=x$, which upon integration gives $x=e^{\lambda r}$ with $-\infty<r<+\infty$. Moreover, the potential function (56) becomes

$$W(x)=-\frac{\lambda^2}{2}\left(A_+e^{2\lambda r}-A_0e^{\lambda r}+A_-\right). \tag{68}$$

This corresponds to the 1D Morse potential $V_{\mathrm{M}}(r)=V_2\,e^{2\lambda r}-V_1\,e^{\lambda r}$ with the following choice of parameters

$$A_+=-2V_2/\lambda^2\,,\quad A_0=-2V_1/\lambda^2\,,\quad A_-=2E/\lambda^2\,, \tag{69}$$

where $\lambda$ is now a physical parameter that gives a measure for the range of the potential. This potential has discrete as well as continuous energy spectra with the former being finite. Thus, we could identify this problem with the solution in subsection 2.2.1 in terms of the continuous dual Hahn polynomial $S_n^\tau\left(z^2;\frac{\nu+1}{2},\frac{\nu+1}{2}\right)$, then using (27) and (A7b) we obtain the following

$$2\alpha=\nu+1\,,\qquad 2\beta=1+\sqrt{1-8V_2/\lambda^2}\,,\qquad \tau=\frac{1}{2}-\frac{2V_1}{\lambda^2}\,,\qquad z^2=2E/\lambda^2\,, \tag{70}$$

with $E\geq 0$ and $V_2\leq\lambda^2/8$. If $V_1\leq\lambda^2/4$ then $\tau>0$ and the spectrum is entirely continuous with the following scattering wavefunction

$$\psi(r,E)=p(E)e^{(\nu+1)\lambda r/2}\exp\left(-\beta e^{\lambda r}\right)\sum\nolimits_{n=0}^{\infty}c_nS_n^\tau\left(\kappa^2/\lambda^2;\tfrac{\nu+1}{2},\tfrac{\nu+1}{2}\right)L_n^\nu(e^{\lambda r})\,, \tag{71}$$

where $\kappa=\sqrt{2E}$ and $p(E)$ is given by Eq. (29). Moreover, the asymptotics ($n\to\infty$) of the continuous dual Hahn polynomial $S_n^\tau\left(z^2;\alpha,\alpha\right)$ gives the scattering phase shift as $\delta(z)=\arg\Gamma(2\mathrm{i}z)-\arg\Gamma(\tau+\mathrm{i}z)-2\arg\Gamma(\alpha+\mathrm{i}z)$ (see, for example, Eq. A17 in Appendix A of [7]), which for this case reads as follows

$$\delta(E)=\arg\Gamma(2\mathrm{i}\kappa/\lambda)-\arg\Gamma\left(\tfrac{1}{2}-\tfrac{2V_1}{\lambda^2}+\mathrm{i}\kappa/\lambda\right)-2\arg\Gamma\left(\tfrac{\nu+1}{2}+\mathrm{i}\kappa/\lambda\right), \tag{72}$$

However, if $V_1>\lambda^2/4$ then $\tau<0$ and the spectrum is a mix of continuous and discrete energies. The latter is a finite set with $z_m^2=-(m+\tau)^2$, which gives the following energy spectrum formula

$$E_m=-\frac{\lambda^2}{2}\left(m+\tfrac{1}{2}-2V_1/\lambda^2\right)^2, \tag{73}$$

where $m=0,1,2,..,N$ and $N$ being the largest integer less than or equal to $-\tau=-\frac{1}{2}+2V_1/\lambda^2$. The complete wavefunction for a given continuous energy $E$ and discrete level $m$ is written using Eq. (30) as follows

$$\psi_m(r,E)=e^{(\nu+1)\lambda r/2}\exp\left(-\beta e^{\lambda r}\right)$$
$$\left\{p(E)\sum\nolimits_{n=0}^{\infty}c_nS_n^\tau\left(\kappa^2/\lambda^2;\tfrac{\nu+1}{2},\tfrac{\nu+1}{2}\right)L_n^\nu(e^{\lambda r})+p(m)\sum\nolimits_{n=0}^{N}c_nS_n^\tau\left(-(m+\tau)^2;\tfrac{\nu+1}{2},\tfrac{\nu+1}{2}\right)L_n^\nu(e^{\lambda r})\right\} \tag{74}$$

where $p(m)$ is given below Eq. (30).

## 4.2 The Jacobi-type differential equation

For the Jacobi-type differential equation (5), we choose the coordinate transformation such that $-1\le x(r)\le +1$ and $(x^2-1)\left(\xi'/\xi\right)=a-b+x\left(a+b\right)$ whose solution is $\xi(x)=(1-x)^a(1+x)^b$, where $a$ and $b$ are dimensionless real parameters. Consequently, Eq. (54) becomes

$$\frac{\lambda^2\xi^2}{1-x^2}\left\{(1-x^2)\frac{d^2}{dx^2}-\left[a-b+x\left(a+b\right)\right]\frac{d}{dx}-\frac{2}{\lambda^2}(1-x)^{1-2a}(1+x)^{1-2b}W(x)\right\}\psi=0. \quad (75)$$

Resulting in the following choice of function

$$W(x)=-\frac{\lambda^2}{2}(1-x)^{2a-1}(1+x)^{2b-1}\left[\frac{A_+}{1+x}+\frac{A_-}{1-x}+A_1x-A_0\right], \quad (76)$$

and turning Schrödinger equation into the differential equation (5) for $y(x)=\psi(r)$. Hence, the Jacobi-type differential equation (5) is equivalent to the Schrödinger equation in quantum mechanics for the interaction potential $V(r)$, which is equal to the function (76) aside from an overall additive constant ($-E$). In the following three subsections, we give examples from this class with $A_1=0$ and corresponding to $(a,b)=\left(1,\tfrac{1}{2}\right)$, $(a,b)=\left(\tfrac{1}{2},\tfrac{1}{2}\right)$, $(a,b)=(1,0)$.

### 4.2.1 The hyperbolic Pöschl-Teller potential $(a,b)=\left(1,\tfrac{1}{2}\right)$:

These values of $a$ and $b$ make $\xi(x)=\left(1-x\right)\sqrt{1+x}$, which upon integration gives $x=2\tanh(\lambda r)^2$ $-1$ and $r\ge 0$. Moreover, the potential function (76) with $A_1=0$ becomes

$$W(x)=-\frac{\lambda^2}{2}\left[\frac{A_+}{\sinh(\lambda r)^2}-\frac{2A_0}{\cosh(\lambda r)^2}+A_-\right]. \quad (77)$$

This corresponds to the hyperbolic Pöschl-Teller potential, which we can write as $V_{\mathrm{PT}}(r)=$ $\frac{1}{4}\left[\frac{A(A-\lambda)}{\sinh^2(\lambda r)}+\frac{\lambda B}{\cosh^2(\lambda r)}\right]$ with the following choice of parameters

$$A_+=-A(A-\lambda)/2\lambda^2,\quad A_0=B/4\lambda,\qquad A_-=2E/\lambda^2, \quad (78)$$

where the physical parameter $\lambda$ is a measure of the range of the potential. This problem is known to have both continuous as well as discrete energy spectra with the latter being finite. Thus, we could associate the solution of this problem with that of subsection 3.2.1 in terms of the Wilson polynomial $W_n(z^2;\sigma+i\tau,\sigma-i\tau,\gamma,\gamma)$, then we obtain from (43) and (B12c) the following

$$\nu=\pm\left(\tfrac{A}{\lambda}-\tfrac{1}{2}\right),\ \gamma=\alpha=\frac{\mu+1}{2},\ \sigma=\beta+\frac{1}{4},\ 2\beta=\begin{cases}A/\lambda\\ 1-A/\lambda\end{cases},\ 2\tau=\sqrt{\frac{B}{\lambda}-\frac{1}{4}},\ z^2=E/\lambda^2, \quad (79)$$

where $E\ge 0$ and the top (bottom) sign corresponds to positive (negative) potential parameter $A$. If $B\ge\lambda/4$ then $\tau^2>0$ and the spectrum is purely continuous. The corresponding scattering wavefunction reads as follows

$$\psi(r,E)_{A>0} = \cosh(\lambda r)^{-\mu-1}\tanh(\lambda r)^{A/\lambda}$$
$$p(E)\sum_{n=0}^{\infty} c_n^+ W_n\left(E/\lambda^2; \tfrac{A}{2\lambda}+\tfrac{1}{4}+\mathrm{i}\tau, \tfrac{A}{2\lambda}+\tfrac{1}{4}-\mathrm{i}\tau, \tfrac{\mu+1}{2}, \tfrac{\mu+1}{2}\right) P_n^{\left(\mu,\frac{A}{\lambda}-\frac{1}{2}\right)}\left(x(r)\right) \tag{80a}$$

$$\psi(r,E)_{A<0} = \cosh(\lambda r)^{-\mu-1}\tanh(\lambda r)^{1-A/\lambda}$$
$$p(E)\sum_{n=0}^{\infty} c_n^- W_n\left(E/\lambda^2; -\tfrac{A}{2\lambda}+\tfrac{3}{4}+\mathrm{i}\tau, -\tfrac{A}{2\lambda}+\tfrac{3}{4}-\mathrm{i}\tau, \tfrac{\mu+1}{2}, \tfrac{\mu+1}{2}\right) P_n^{\left(\mu,\frac{1}{2}-\frac{A}{\lambda}\right)}\left(x(r)\right) \tag{80b}$$

where $p(E)$ is given by Eq. (45) and $c_n^{\pm} = c_n(\pm\nu > 0)$. Moreover, the asymptotics ($n\to\infty$) of the Wilson polynomial $W_n(z^2;\sigma+\mathrm{i}\tau,\sigma-\mathrm{i}\tau,\gamma,\gamma)$ gives the scattering phase shift as (see Ref. [24] and Eq. B12 in Appendix B of [8])

$$\delta(z) = \arg\Gamma(2\mathrm{i}z) - \arg\Gamma\left[\sigma+\mathrm{i}(z+\tau)\right] - \arg\Gamma\left[\sigma+\mathrm{i}(z-\tau)\right] - 2\arg\Gamma(\gamma+\mathrm{i}z), \tag{81}$$

which for this case reads as follows

$$\delta(E) = \arg\Gamma\left(2\mathrm{i}\tfrac{\kappa}{\lambda}\right) - \arg\Gamma\left[\tfrac{\nu+1}{2}+\mathrm{i}\left(\tfrac{\kappa}{\lambda}+\tau\right)\right] - \arg\Gamma\left[\tfrac{\nu+1}{2}+\mathrm{i}\left(\tfrac{\kappa}{\lambda}-\tau\right)\right] - 2\arg\Gamma\left(\tfrac{\mu+1}{2}+\mathrm{i}\tfrac{\kappa}{\lambda}\right), \tag{82}$$

with $\kappa=\sqrt{E}$. On the other hand, if $B<\lambda/4$ then $\tau^2<0$ and the spectrum is a mix of continuous and discrete energies. The latter is a finite set with $z_m^2 = -\left(m+\sigma-\sqrt{-\tau^2}\right)^2$, which gives the following energy spectrum formula

$$E_m = -\frac{\lambda^2}{4}\left(2m+\nu+1-\sqrt{\tfrac{1}{4}-\tfrac{B}{\lambda}}\right)^2, \tag{83}$$

where $m=0,1,2,..,N$ and $N$ being the largest integer less than or equal to $\frac{1}{2}\sqrt{\frac{1}{4}-\frac{B}{\lambda}}-\frac{\nu+1}{2}$. The complete wavefunction for a given continuous energy $E$ and discrete level $m$ is written using Eq. (46) as follows

$$\psi_m(r,E) = \cosh(\lambda r)^{-\mu-1}\tanh(\lambda r)^{\nu+\frac{1}{2}} \times$$
$$\left\{p(E)\sum_{n=0}^{\infty} c_n W_n\left(E/\lambda^2; \tfrac{\nu+1}{2}+\mathrm{i}\tau, \tfrac{\nu+1}{2}-\mathrm{i}\tau, \tfrac{\mu+1}{2}, \tfrac{\mu+1}{2}\right) P_n^{(\mu,\nu)}\left(x(r)\right) + \right.$$
$$\left. p(m)\sum_{n=0}^{N} c_n W_n\left(-E_m/\lambda^2; \tfrac{\nu+1}{2}+\mathrm{i}\tau, \tfrac{\nu+1}{2}-\mathrm{i}\tau, \tfrac{\mu+1}{2}, \tfrac{\mu+1}{2}\right) P_n^{(\mu,\nu)}\left(x(r)\right)\right\} \tag{84}$$

with $\nu = \pm\left(\frac{A}{\lambda}-\frac{1}{2}\right)$ for $\pm A>0$ and $p(m)$ is given below Eq. (46).

### 4.2.2 The trigonometric Scarf potential $(a,b)=\left(\frac{1}{2},\frac{1}{2}\right)$:

These values of $a$ and $b$ give $\xi(x)=\sqrt{1-x^2}$, which upon integration yields $x(r)=-\cos\left(\pi r/L\right)$ and $0\le r\le L$. Moreover, the potential function (76) with $A_1=0$ becomes

$$W(x) = -\frac{\lambda^2}{2}\left[\frac{(A_+ + A_-)+(A_+ - A_-)\cos(\pi r/L)}{\sin^2(\pi r/L)} - A_0\right], \tag{85}$$

This corresponds to the trigonometric Scarf potential, which we can write as

$$V_{\rm Sc}(r)=\frac{(A^2+B^2-\lambda A)-B(2A-\lambda)\cos(\lambda r)}{2\sin^2(\lambda r)}, \tag{86}$$

with $\lambda=\pi/L$ and giving the following parameter map

$$2A_+=\tfrac{1}{4}-\left(\tfrac{A}{\lambda}-\tfrac{B}{\lambda}-\tfrac{1}{2}\right)^2,\qquad 2A_-=\tfrac{1}{4}-\left(\tfrac{A}{\lambda}+\tfrac{B}{\lambda}-\tfrac{1}{2}\right)^2,\qquad A_0=-2E/\lambda^2. \tag{87}$$

Since this is a confining potential box, we expect that the spectrum is entirely discrete and its size is infinite. Therefore, we cannot associate this problem with the discrete Racah polynomial since the size of its spectrum is fixed at a finite integer *N*. On the other hand, we can associate the problem with the Wilson polynomial for two reasons: (1) it has a discrete part, and (2) we can impose conditions to extend the spectrum size to infinity. Hence, associating the problem with $W_n(z^2;\sigma+i\tau,\sigma-i\tau,\gamma,\gamma)$ and using Eq. (43) and (B12c), we obtain

$$\nu=\pm\left(\tfrac{A}{\lambda}-\tfrac{B}{\lambda}-\tfrac{1}{2}\right),\; 2\beta=\nu+\tfrac{1}{2}=\begin{cases}\frac{A}{\lambda}-\frac{B}{\lambda}\\ 1-\frac{A}{\lambda}+\frac{B}{\lambda}\end{cases},\; 2\alpha=\mu+\tfrac{1}{2},\; 2\sigma=\nu+1,\; 2\gamma=\mu+1, \tag{88a}$$

$$\tau=\sqrt{-2E/\lambda^2},\quad z^2=-\tfrac{1}{4}\left(\tfrac{A}{\lambda}+\tfrac{B}{\lambda}-\tfrac{1}{2}\right)^2. \tag{88b}$$

where the top (bottom) sign in (88a) corresponds to the case where the potential parameter *A* is greater (less) than *B*. The Wilson polynomial spectrum formula $z_m^2=-\left(m+\sigma-\sqrt{-\tau^2}\right)^2$ gives the following bound states energies

$$E_m=\frac{\lambda^2}{2}\left[m+\tfrac{\nu+1}{2}+\tfrac{1}{2}\left(\tfrac{A}{\lambda}+\tfrac{B}{\lambda}-\tfrac{1}{2}\right)\right]^2=\frac{\lambda^2}{2}\begin{cases}\left(m+A/\lambda\right)^2 & ,A>B\\ \left(m+\tfrac{1}{2}+B/\lambda\right)^2 & ,B>A\end{cases} \tag{89}$$

Now, the spectrum size constraint states that *N* must be the largest integer less than or equal to $\sqrt{-\tau^2}-\sigma$, which is satisfied by any value of *N* up to infinity. The bound state wavefunction for a given level *m* is written using Eq. (46) as follows

$$\psi_m(r)=p(m)\left(\cos(\pi r/2L)\right)^{\mu+1/2}\left(\sin(\pi r/2L)\right)^{\nu+1/2}\times$$
$$\sum_{n=0}^{\infty}c_nW_n\left(-z_m^2;\tfrac{\nu+1}{2}-\tfrac{\kappa_m}{\lambda},\tfrac{\nu+1}{2}+\tfrac{\kappa_m}{\lambda},\tfrac{\mu+1}{2},\tfrac{\mu+1}{2}\right)P_n^{(\mu,\nu)}\left(-\cos(\pi r/L)\right) \tag{90}$$

where $\kappa_m=\sqrt{2E_m}$ and $\nu=\pm\left(\tfrac{A}{\lambda}-\tfrac{B}{\lambda}-\tfrac{1}{2}\right)$ for $\pm(A-B)>0$.

### 4.2.3 The hyperbolic Eckart potential $(a,b)=(1,0)$:

These values of *a* and *b* give $\xi(x)=(1-x)$, which upon integration yields $x(r)=1-2e^{-\lambda r}$ and $r\geq 0$. Moreover, the potential function (76) with $A_1=0$ becomes

$$W(x)=\frac{-\lambda^2/2}{1-e^{-\lambda r}}\left[\frac{A_-}{2}-A_0+\frac{A_+/2}{e^{\lambda r}-1}\right]-\frac{\lambda^2}{2}A_0. \tag{91}$$

This corresponds to the hyperbolic Eckart potential, which we can write as

$$V_{\rm Ek}(r)=\frac{1/2}{1-e^{-\lambda r}}\left[\lambda B+\frac{A(A-\lambda)}{e^{\lambda r}-1}\right]=\frac{1}{4}\left[\frac{A(A-\lambda)/2}{\sinh^2(\lambda r/2)}+\frac{\lambda B}{\tanh(\lambda r/2)}+\lambda B\right]. \tag{92}$$

Comparing this with (91) gives the following parameters map

$$A_+=-2\tfrac{A}{\lambda}\left(\tfrac{A}{\lambda}-1\right),\qquad A_0=2E/\lambda^2\,,\qquad A_-=2\left(2E-\lambda B\right)/\lambda^2 \tag{93}$$

This potential has an energy spectrum made up of a continuous part and a finite discrete part. Thus, we may associate the solution of this problem with that of subsection 3.2.1 in terms of the Wilson polynomial $W_n(z^2;\sigma+i\tau,\sigma-i\tau,\gamma,\gamma)$. Hence, using (43) and (B12c) we obtain the following polynomial parameters assignment

$$\nu=\pm\left(2\tfrac{A}{\lambda}-1\right),\ \gamma=\alpha=\frac{\mu+1}{2},\ \sigma=\beta=\begin{cases}A/\lambda\\ 1-A/\lambda\end{cases},\ \tau=\kappa/\lambda\,,\ z^2=\left(\kappa/\lambda\right)^2-B/\lambda\,, \tag{94}$$

where $\kappa=\sqrt{2E}$ and the top (bottom) sign corresponds to the case where $A$ is positive (negative), respectively. If the energy $E$ is positive and also $E\geq\lambda B/2$ then $z^2\geq 0$ and $\tau^2>0$ making the spectrum purely continuous. The corresponding scattering wave function becomes

$$\psi(r,E)_{A>0}=p(E)e^{-(\mu+1)\lambda r/2}\left(1-e^{-\lambda r}\right)^{A/\lambda}$$
$$\sum\nolimits_{n=0}^{\infty}c_n^+W_n\left(z^2;\tfrac{A}{\lambda}+i\tfrac{\kappa}{\lambda},\tfrac{A}{\lambda}-i\tfrac{\kappa}{\lambda},\tfrac{\mu+1}{2},\tfrac{\mu+1}{2}\right)P_n^{\left(\mu,2\frac{A}{\lambda}-1\right)}\left(1-2e^{-\lambda r}\right) \tag{95a}$$

$$\psi(r,E)_{A<0}=p(E)e^{-(\mu+1)\lambda r/2}\left(1-e^{-\lambda r}\right)^{1-A/\lambda}$$
$$\sum\nolimits_{n=0}^{\infty}c_n^-W_n\left(z^2;1-\tfrac{A}{\lambda}+i\tfrac{\kappa}{\lambda},1-\tfrac{A}{\lambda}-i\tfrac{\kappa}{\lambda},\tfrac{\mu+1}{2},\tfrac{\mu+1}{2}\right)P_n^{\left(\mu,1-2\frac{A}{\lambda}\right)}\left(1-2e^{-\lambda r}\right) \tag{95b}$$

where $p(E)$ is given by Eq. (45) and $c_n^\pm=c_n(\pm\nu>0)$. Moreover, the asymptotics ($n\to\infty$) of the Wilson polynomial gives the following scattering phase shift

$$\delta(E)=\arg\Gamma\left(2iz\right)-\arg\Gamma\left[\tfrac{\nu+1}{2}+i\left(z+\tfrac{\kappa}{\lambda}\right)\right]-\arg\Gamma\left[\tfrac{\nu+1}{2}+i\left(z-\tfrac{\kappa}{\lambda}\right)\right]-2\arg\Gamma\left(\tfrac{\mu+1}{2}+iz\right), \tag{96}$$

where $z=\sqrt{\left(\kappa/\lambda\right)^2-B/\lambda}$ and $\nu=\pm\left(2\tfrac{A}{\lambda}-1\right)$ for $\pm A>0$. On the other hand, for states with a mix of continuous and discrete energies, the finite energy spectrum is obtained from the spectrum formula of the Wilson polynomial that reads $z_m^2=-\left(m+\sigma-\sqrt{-\tau^2}\right)^2$. Consequently, we obtain the following bound state energies

$$E_m=-\frac{\lambda^2}{8}\left(m+\tfrac{\nu+1}{2}-\frac{B/\lambda}{m+\frac{\nu+1}{2}}\right)^2. \tag{97}$$

where $m = 0,1,2,..,N$ . The size of the energy spectrum is obtained from the condition that $z^2 < 0$ (i.e. $E < \lambda B/2$ ), which makes $N$ the largest integer less than or equal to $-\sigma + \sqrt{-B/\lambda}$ . The complete wavefunction for a given continuous energy $E$ and discrete level $m$ is written using Eq. (46) as follows

$$\begin{gathered} \psi_m(r,E) = e^{-(\mu+1)\lambda r/2}\left(1-e^{-\lambda r}\right)^{\frac{\nu+1}{2}} \times \\ \Big\{ p(E)\sum_{n=0}^{\infty} c_n W_n\left(z^2;\tfrac{\nu+1}{2}+\mathrm{i}\,\tau,\tfrac{\nu+1}{2}-\mathrm{i}\,\tau,\tfrac{\mu+1}{2},\tfrac{\mu+1}{2}\right) P_n^{(\mu,\nu)}\left(1-2e^{-\lambda r}\right) + \\ p(m)\sum_{n=0}^{N} c_n W_n\left(z_m^2;\tfrac{\nu+1}{2}+\mathrm{i}\,\tau,\tfrac{\nu+1}{2}-\mathrm{i}\,\tau,\tfrac{\mu+1}{2},\tfrac{\mu+1}{2}\right) P_n^{(\mu,\nu)}\left(1-2e^{-\lambda r}\right)\Big\} \end{gathered} \tag{98}$$

with $\nu = \pm\left(2\frac{A}{\lambda}-1\right)$ for $\pm A > 0$, $z^2 = \left(2E/\lambda^2\right) - B/\lambda$, $z_m^2 = -\left(m+\sigma-\sqrt{-2E_m/\lambda^2}\right)^2$ and $p(m)$ is given below Eq. (46).

# 5. Conclusion

We introduced two ordinary second order linear differential equations of the Laguerre- and Jacobi-type and derived their solutions as infinite series of square integrable functions written in terms of the Laguerre and Jacobi polynomials. The expansion coefficients in the series are orthogonal polynomials satisfying $ST^2R^2$s. Some of these polynomials are new and were not treated previously in the mathematics literature while most are well-known hypergeometric type orthogonal polynomials such as the Meixner-Pollaczek, continuous dual Hahn and Wilson polynomials. Physical applications of the solutions of these differential equations in quantum mechanics were also presented where we have given the corresponding interaction potentials and associated phase shift for the scattering states and energy spectrum for the bound states. In this work, we obtain only one solution of the differential equation. It turned out that this solution meets all physical requirements including the boundary conditions for both the discrete bound states as well as the continuum scattering states. We did not search for a second independent solution, which might be written in terms of another kind of independent polynomials such as the polynomials of the second kind and/or the associated polynomials.

The details of the solution of the Jacobi-type differential equation (5) with $A_1 \neq 0$ could not be given due to the lack of knowledge of the analytic properties of the corresponding orthogonal polynomials $H_n^{(\mu,\nu)}(z^{-1};\theta,\sigma)$ and $G_n^{(\mu,\nu)}(k;\tau,\sigma)$, which are defined up to now by their $ST^2R^2$s shown in subsection 3.1. Due to the prime significance of these polynomials in physics, we call upon experts in the field to derive their properties (weight function, asymptotics, generating function, etc.).

## Appendix A: The Laguerre class $ST^2R^2$

The requirement that the substitution of $y(x)$ given by (2) in the differential equations (1) result in a $ST^2R^2$ for the expansion coefficients $\{f_n\}$ means that the action of the differential operator (1) on the basis element (4) gives three terms containing $L_n^\nu(x)$ and $L_{n\pm1}^\nu(x)$ with constant multiplicative factors. Now, this differential operator action on $\phi_n(x)$ becomes

$$c_n x^\alpha e^{-\beta x}\left\{x\frac{d^2}{dx^2}+\left[a+2\alpha+x\left(b-2\beta\right)\right]\frac{d}{dx}+x\left[A_++\beta\left(\beta-b\right)\right]\right.$$
$$\left.+\frac{A_-+\alpha\left(\alpha+a-1\right)}{x}-\left[A_0+\beta\left(2\alpha+a\right)-b\alpha\right]\right\}L_n^\nu(x) \tag{A1}$$

Using the differential equation of the Laguerre polynomials, $\left[x\frac{d^2}{dx^2}+\left(\nu+1-x\right)\frac{d}{dx}+n\right]L_n^\nu(x)=$ 0, turns (A1) into the following

$$c_n x^\alpha e^{-\beta x}\left\{\left[a+2\alpha-\nu-1+x\left(b-2\beta+1\right)\right]\frac{d}{dx}+x\left[A_++\beta\left(\beta-b\right)\right]\right.$$
$$\left.+\frac{A_-+\alpha\left(\alpha+a-1\right)}{x}-\left[A_0+\beta\left(2\alpha+a\right)-b\alpha+n\right]\right\}L_n^\nu(x) \tag{A2}$$

The differential property of the Laguerre polynomials, $x\frac{d}{dx}L_n^\nu(x)=nL_n^\nu(x)-\left(n+\nu\right)L_{n-1}^\nu(x)$, turns (A2) into the following

$$\left\{\frac{A_-+\alpha\left(\alpha+a-1\right)+n\left(a+2\alpha-\nu-1\right)}{x}-\left[A_0+\beta\left(2\alpha+a\right)-b\alpha+n\left(2\beta-b\right)\right]\right.$$
$$\left.+x\left[A_++\beta\left(\beta-b\right)\right]\right\}L_n^\nu(x)-\left(n+\nu\right)\left[\frac{a+2\alpha-\nu-1}{x}+b-2\beta+1\right]L_{n-1}^\nu(x) \tag{A3}$$

where we have hidden the overall multiplicative factor $c_n x^\alpha e^{-\beta x}$. The three-term recursion requirement on (A3) together with the recursion relation of the Laguerre polynomials, $xL_n^\nu(x)=$ $\left(2n+\nu+1\right)L_n^\nu(x)-\left(n+\nu\right)L_{n-1}^\nu(x)-\left(n+1\right)L_{n-1}^\nu(x)$, allow only constant terms multiplying $L_{n-1}^\nu(x)$ in the second line of (A3) and only linear terms in $x$ multiplying $L_n^\nu(x)$. These constraints lead to the following two set of conditions:

(a) $2\alpha=\nu+1-a$, $\quad \nu^2=(1-a)^2-4A_-$. (A4a)

(b) $2\beta=b+1$, $\quad b^2=1+4A_+$. (A4b)

The second set of conditions (A4b) are obtained by multiplying (A3) by $x$ (i.e., factoring out $x^{-1}$ making the overall multiplicative factor $c_n x^{\alpha-1}e^{-\beta x}$) then imposing the three-term requirement. Reality requires that $A_-\le\left(\frac{1-a}{2}\right)^2$ for case (A4a) and that $A_+\ge-\frac{1}{4}$ for case (A4b). Applying the two sets of constraints together with the three-term recursion relation of the Laguerre polynomials, we get

$$\begin{gathered}\left[(2n+\nu+1)\left(\gamma-\tfrac{1}{2}\right)-A_0-\tfrac{ab}{2}\right]L_n^\nu(x)\\ -\frac{1}{2}(n+\nu)(2\gamma+b+1-2\beta)L_{n-1}^\nu(x)-\frac{1}{2}(n+1)(2\gamma+2\beta-b-1)L_{n+1}^\nu(x)\end{gathered} \tag{A5a}$$

$$\begin{gathered}\left[-(2n+\nu+1)(n+\omega)+A_-+\tfrac{1}{2}(\nu+1)^2-\tfrac{1}{2}(\nu+1)(2\alpha+a)+\alpha(\alpha+a-1)\right]L_n^\nu(x)\\ +(n+\nu)\left[n+\omega+\tfrac{1}{2}(\nu+1-a-2\alpha)\right]L_{n-1}^\nu(x)+(n+1)\left[n+\omega-\tfrac{1}{2}(\nu+1-a-2\alpha)\right]L_{n+1}^\nu(x)\end{gathered} \tag{A5b}$$

where $\gamma=A_+ +\beta(\beta-b-1)+\frac{1+b}{2}$ and $\omega=A_0+\frac{1}{2}(\nu+ab+1)$. Rewriting these as three-term recursion relation for $\phi_n(x)$ instead of $L_n^\nu(x)$ (i.e., taking care of the normalization factor $c_n$) gives

$$\begin{gathered}\left[(2n+\nu+1)\left(\gamma-\tfrac{1}{2}\right)-A_0-\tfrac{ab}{2}\right]\phi_n(x)\\ -\frac{1}{2}\sqrt{n(n+\nu)}(2\gamma+b+1-2\beta)\phi_{n-1}(x)-\frac{1}{2}\sqrt{(n+1)(n+\nu+1)}(2\gamma+2\beta-b-1)\phi_{n+1}(x)\end{gathered} \tag{A6a}$$

$$\begin{gathered}\left[-(2n+\nu+1)(n+\omega)+A_-+\tfrac{1}{2}(\nu+1)^2-\tfrac{1}{2}(\nu+1)(2\alpha+a)+\alpha(\alpha+a-1)\right]\varphi_n(x)\\ +\sqrt{n(n+\nu)}\left[n+\omega+\tfrac{1}{2}(\nu+1-a-2\alpha)\right]\varphi_{n-1}(x)+\sqrt{(n+1)(n+\nu)}\left[n+\omega-\tfrac{1}{2}(\nu+1-a-2\alpha)\right]\varphi_{n+1}(x)\end{gathered} \tag{A6b}$$

where $\varphi_n(x)=x^{-1}\phi_n(x)$. These three-term recursion relations have the generic form $s_n\phi_n(x)+r_n\phi_{n-1}(x)+t_n\phi_{n+1}(x)$. Requiring it to be a symmetric three-term recursion relation dictates that $r_n=t_{n-1}$. For the case (A6a), this requires that $2\beta=b+1$, whereas for the case (A6b), we must have $2\alpha=\nu+2-a$. Putting all of the above findings together, we obtain the following two alternative scenarios for the solution of the Laguerre-type differential equation

(a) $2\alpha=\nu+1-a$, $\quad 2\beta=b+1$, $\quad \nu^2=(1-a)^2-4A_-$, $\quad A_-\le\left(\frac{1-a}{2}\right)^2$. (A7a)

(b) $2\alpha=\nu+2-a$, $\quad 2\beta=b+1$, $\quad b^2=1+4A_+$, $\quad A_+\ge-\frac{1}{4}$. (A7b)

together with the corresponding ST²R² for the expansion coefficients $\{f_n\}$

$$\begin{gathered}\left[(2n+\nu+1)\left(A_+-\tfrac{1}{4}b^2-\tfrac{1}{4}\right)-A_0-\tfrac{ab}{2}\right]f_n(z)\\ -\left(A_+-\tfrac{1}{4}b^2+\tfrac{1}{4}\right)\left[\sqrt{n(n+\nu)}f_{n-1}(z)+\sqrt{(n+1)(n+\nu+1)}f_{n+1}(z)\right]=0\end{gathered} \tag{A8a}$$

$$\begin{gathered}\left[-(2n+\nu+1)\left(n+A_0+\tfrac{\nu+ab+1}{2}\right)+A_-+\tfrac{1}{4}\left(\nu^2-1\right)-\tfrac{1}{4}(a-1)^2\right]f_n(z)\\ +\left(n+A_0+\tfrac{\nu+ab}{2}\right)\sqrt{n(n+\nu)}f_{n-1}(z)+\left(n+1+A_0+\tfrac{\nu+ab}{2}\right)\sqrt{(n+1)(n+\nu+1)}f_{n+1}(z)=0\end{gathered} \tag{A8b}$$

Taking $f_0(z)=1$ these ST$^2$R$^2$ give $f_n(z)$ as a polynomial of degree $n$ in $z$, where $z \propto \left(A_0+\frac{ab}{2}\right)\left(A_+-\frac{1}{4}b^2+\frac{1}{4}\right)^{-1}$ in (A8a) and $z \propto A_-+\frac{1}{4}\left(\nu^2-1\right)-\frac{1}{4}(a-1)^2$ in (A8b). These solutions are obtained in section 2 in terms of orthogonal polynomials with continuous and discrete spectra.

## Appendix B: The Jacobi class ST$^2$R$^2$

The requirement that the substitution of $y(x)$ given by (6) in the differential equations (5) result in a ST$^2$R$^2$ for the expansion coefficients $\{f_n\}$ means that the action of the differential operator (5) on the basis element (6) must give three terms containing $P_n^{(\mu,\nu)}(x)$ and $P_{n\pm 1}^{(\mu,\nu)}(x)$ with constant multiplicative factors. Now, this differential operator action on $\phi_n(x)$ becomes

$$c_n(1-x)^{\alpha}(1+x)^{\beta}\left\{(1-x^2)\frac{d^2}{dx^2}-\left[2\alpha+a-2\beta-b+x\left(2\alpha+a+2\beta+b\right)\right]\frac{d}{dx}+A_1x\right.$$
$$\left.+\frac{A_-+2\alpha\left(\alpha+a-1\right)}{1-x}+\frac{A_++2\beta\left(\beta+b-1\right)}{1+x}-\left[A_0+\left(\alpha+\beta\right)\left(\alpha+\beta+a+b-1\right)\right]\right\}P_n^{(\mu,\nu)}(x) \tag{B1}$$

Using the differential equation of the Jacobi polynomials,

$$\left\{\left(1-x^2\right)\frac{d^2}{dx^2}-\left[\left(\mu+\nu+2\right)x+\mu-\nu\right]\frac{d}{dx}+n\left(n+\mu+\nu+1\right)\right\}P_n^{(\mu,\nu)}(x)=0, \tag{B2}$$

turns (B1) into the following

$$c_n(1-x)^{\alpha}(1+x)^{\beta}\left\{-\left[\left(2\alpha+a-\mu-2\beta-b+\nu\right)+x\left(2\alpha+a-\mu+2\beta+b-\nu-2\right)\right]\frac{d}{dx}+A_1x\right.$$
$$\left.+\frac{A_-+2\alpha\left(\alpha+a-1\right)}{1-x}+\frac{A_++2\beta\left(\beta+b-1\right)}{1+x}-\left[A_0+\left(\alpha+\beta\right)\left(\alpha+\beta+a+b-1\right)+n\left(n+\mu+\nu+1\right)\right]\right\}P_n^{(\mu,\nu)}(x) \tag{B3}$$

The differential property of the Jacobi polynomials,

$$\left(1-x^2\right)\frac{d}{dx}P_n^{(\mu,\nu)}(x)=-n\left(x+\frac{\nu-\mu}{2n+\mu+\nu}\right)P_n^{(\mu,\nu)}(x)+2\frac{(n+\mu)(n+\nu)}{2n+\mu+\nu}P_{n-1}^{(\mu,\nu)}(x), \tag{B4}$$

turns (B3) into the following

$$\left[\frac{2\alpha+a-\mu-1}{1-x}-\frac{2\beta+b-\nu-1}{1+x}\right]\left[n\left(x+\frac{\nu-\mu}{2n+\mu+\nu}\right)P_n^{(\mu,\nu)}(x)-2\frac{(n+\mu)(n+\nu)}{2n+\mu+\nu}P_{n-1}^{(\mu,\nu)}(x)\right]$$
$$+\left\{\frac{A_-+2\alpha\left(\alpha+a-1\right)}{1-x}+\frac{A_++2\beta\left(\beta+b-1\right)}{1+x}+A_1x-\left[A_0+\left(\alpha+\beta\right)\left(\alpha+\beta+a+b-1\right)+n\left(n+\mu+\nu+1\right)\right]\right\}P_n^{(\mu,\nu)}(x) \tag{B5}$$

where we have hidden the overall multiplicative factor $c_n(1-x)^{\alpha}(1+x)^{\beta}$. The three-term recursion requirement on (B5) together with the recursion relation of the Jacobi polynomials, which reads

$$xP_n^{(\mu,\nu)}(x) = \frac{\nu^2-\mu^2}{(2n+\mu+\nu)(2n+\mu+\nu+2)} P_n^{(\mu,\nu)}(x) + \frac{2(n+\mu)(n+\nu)}{(2n+\mu+\nu)(2n+\mu+\nu+1)} P_{n-1}^{(\mu,\nu)}(x) + \frac{2(n+1)(n+\mu+\nu+1)}{(2n+\mu+\nu+1)(2n+\mu+\nu+2)} P_{n+1}^{(\mu,\nu)}(x) \tag{B6}$$

allow only constant terms multiplying $P_{n-1}^{(\mu,\nu)}(x)$ in the first line of (B5) and only linear terms in $x$ multiplying $P_n^{(\mu,\nu)}(x)$. These constraints lead to the following three set of conditions:

(a) $2\alpha = \mu + 1 - a$, $2\beta = \nu + 1 - b$, $\mu^2 = (1-a)^2 - 2A_-$, $\nu^2 = (1-b)^2 - 2A_+$, $2A_- \le (1-a)^2$, $2A_+ \le (1-b)^2$. (B7a)

(b) $A_1 = 0$, $2\alpha = \mu + 1 - a$, $\mu^2 = (1-a)^2 - 2A_-$, $2A_- \le (1-a)^2$. (B7b)

(c) $A_1 = 0$, $2\beta = \nu + 1 - b$, $\nu^2 = (1-b)^2 - 2A_+$, $2A_+ \le (1-b)^2$. (B7c)

The second set of conditions (B7b) are obtained by multiplying (B5) by $1+x$ then imposing the three-term requirement. That is, factoring out $\frac{1}{1+x}$ making the overall factor multiplying (B5) equal to $c_n(1-x)^\alpha(1+x)^{\beta-1}$. The third set of conditions (B7c) are obtained by multiplying (B5) by $1-x$. That is, factoring out $\frac{1}{1-x}$ making the overall multiplicative factor $c_n(1-x)^{\alpha-1}(1+x)^\beta$. Reality requires that $2A_- \le (1-a)^2$ for cases (B7a) and (B7b) whereas $2A_+ \le (1-b)^2$ for cases (B7a) and (B7c). Applying the above three sets of constraints together with the three-term recursion relation of the Jacobi polynomials, we get

$$\left[\frac{(\nu^2-\mu^2)A_1}{(2n+\mu+\nu)(2n+\mu+\nu+2)} - A_0 - \frac{1}{4}(2n+\mu+\nu+1)^2 + \frac{1}{4}(a+b-1)^2\right] P_n^{(\mu,\nu)}(x) + \frac{2(n+\mu)(n+\nu)A_1}{(2n+\mu+\nu)(2n+\mu+\nu+1)} P_{n-1}^{(\mu,\nu)}(x) + \frac{2(n+1)(n+\mu+\nu+1)A_1}{(2n+\mu+\nu+1)(2n+\mu+\nu+2)} P_{n+1}^{(\mu,\nu)}(x) \tag{B8a}$$

$$\left\{\frac{2n(n+\mu+\nu+1)(\mu-\nu)\omega}{(2n+\mu+\nu)(2n+\mu+\nu+2)} - \left[\gamma + n(n+\mu+\nu+1)\right]\left[1 + \frac{\nu^2-\mu^2}{(2n+\mu+\nu)(2n+\mu+\nu+1)}\right] + A_+ + 2\beta(\beta+b-1)\right\} P_n^{(\mu,\nu)}(x) - \frac{2(n+\mu)(n+\nu)}{(2n+\mu+\nu)(2n+\mu+\nu+1)}\left[\gamma + (n-\omega)(n+\mu+\nu+1)\right] P_{n-1}^{(\mu,\nu)}(x) - \frac{2(n+1)(n+\mu+\nu+1)}{(2n+\mu+\nu+1)(2n+\mu+\nu+2)}\left[\gamma + n(n+\mu+\nu+1+\omega)\right] P_{n+1}^{(\mu,\nu)}(x) \tag{B8b}$$

$$\left\{\frac{2n(n+\mu+\nu+1)(\nu-\mu)\eta}{(2n+\mu+\nu)(2n+\mu+\nu+2)} - \left[\gamma + n(n+\mu+\nu+1)\right]\left[1 + \frac{\mu^2-\nu^2}{(2n+\mu+\nu)(2n+\mu+\nu+1)}\right] + A_- + 2\alpha(\alpha+a-1)\right\} P_n^{(\mu,\nu)}(x) + \frac{2(n+\mu)(n+\nu)}{(2n+\mu+\nu)(2n+\mu+\nu+1)}\left[\gamma + (n-\eta)(n+\mu+\nu+1)\right] P_{n-1}^{(\mu,\nu)}(x) + \frac{2(n+1)(n+\mu+\nu+1)}{(2n+\mu+\nu+1)(2n+\mu+\nu+2)}\left[\gamma + n(n+\mu+\nu+1+\eta)\right] P_{n+1}^{(\mu,\nu)}(x) \tag{B8c}$$

where $\gamma = A_0 + (\alpha+\beta)(\alpha+\beta+a+b-1)$, $\omega = 2\beta + b - \nu - 1$ and $\eta = 2\alpha + a - \mu - 1$. Rewriting these as three-term recursion relation for $\phi_n(x)$ instead of $P_n^{(\mu,\nu)}(x)$ (i.e., taking care of the normalization factor $c_n$) gives

$$\left[A_1 C_n - \frac{1}{4}(2n+\mu+\nu+1)^2 + \frac{1}{4}(a+b-1)^2 - A_0\right]\phi_n(x) + A_1\left[D_{n-1}\phi_{n-1}(x) + D_n\phi_{n+1}(x)\right] \tag{B9a}$$

$$\begin{gathered}\left\{\frac{2n(n+\mu+\nu+1)(\mu-\nu)\omega}{(2n+\mu+\nu)(2n+\mu+\nu+2)} - (C_n+1)\left[\gamma+n(n+\mu+\nu+1)\right] + A_+ + 2\beta(\beta+b-1)\right\}\varphi_n^+(x) \\ -D_{n-1}\left[\gamma+(n-\omega)(n+\mu+\nu+1)\right]\varphi_{n-1}^+(x) - D_n\left[\gamma+n(n+\mu+\nu+1+\omega)\right]\varphi_{n+1}^+(x)\end{gathered} \tag{B9b}$$

$$\begin{gathered}\left\{\frac{2n(n+\mu+\nu+1)(\nu-\mu)\eta}{(2n+\mu+\nu)(2n+\mu+\nu+2)} + (C_n-1)\left[\gamma+n(n+\mu+\nu+1)\right] + A_- + 2\alpha(\alpha+a-1)\right\}\varphi_n^-(x) \\ +D_{n-1}\left[\gamma+(n-\eta)(n+\mu+\nu+1)\right]\varphi_{n-1}^-(x) + D_n\left[\gamma+n(n+\mu+\nu+1+\eta)\right]\varphi_{n+1}^-(x)\end{gathered} \tag{B9c}$$

where

$$C_n = \frac{\nu^2-\mu^2}{(2n+\mu+\nu)(2n+\mu+\nu+2)},\quad D_n = \frac{2}{2n+\mu+\nu+2}\sqrt{\frac{(n+1)(n+\mu+1)(n+\nu+1)(n+\mu+\nu+1)}{(2n+\mu+\nu+1)(2n+\mu+\nu+3)}}, \tag{B10}$$

and $\varphi_n^\pm(x) = (1\pm x)^{-1}\phi_n(x)$. These three-term recursion relations have the generic form $s_n\phi_n(x) + r_n\phi_{n-1}(x) + t_n\phi_{n+1}(x)$. Only relation (B9a) is symmetric since $r_n = t_{n-1}$. Requiring the other two to be symmetric dictates that $\omega=1$ (i.e., $2\beta=\nu+2-b$) for (B9b) whereas for case (B9c), we must have $\eta=1$ (i.e., $2\alpha=\mu+2-a$). We also employ the following identities

$$\frac{2n(n+\mu+\nu+1)(\mu-\nu)}{(2n+\mu+\nu)(2n+\mu+\nu+2)} = \frac{2n(n+\mu)}{2n+\mu+\nu} - n(C_n+1), \tag{B11b}$$

$$\frac{2n(n+\mu+\nu+1)(\nu-\mu)}{(2n+\mu+\nu)(2n+\mu+\nu+2)} = \frac{2n(n+\nu)}{2n+\mu+\nu} + n(C_n-1). \tag{B11c}$$

Putting all above findings together, we obtain the following three alternative scenarios for the solution of the Jacobi-type differential equation

(a) $2\alpha=\mu+1-a$, $2\beta=\nu+1-b$, $\mu^2=(1-a)^2-2A_-$, $\nu^2=(1-b)^2-2A_+$, $2A_-\le(1-a)^2$, $2A_+\le(1-b)^2$. (B12a)

(b) $A_1=0$, $2\alpha=\mu+1-a$, $2\beta=\nu+2-b$, $\mu^2=(1-a)^2-2A_-$, $2A_-\le(1-a)^2$. (B12b)

(c) $A_1=0$, $2\alpha=\mu+2-a$, $2\beta=\nu+1-b$, $\nu^2=(1-b)^2-2A_+$, $2A_+\le(1-b)^2$ (B12c)

together with the corresponding $ST^2R^2$ for the expansion coefficients $\{f_n\}$

$$\left[A_1 C_n - \frac{1}{4}(2n+\mu+\nu+1)^2 + \frac{1}{4}(a+b-1)^2 - A_0\right]f_n(x) + A_1\left[D_{n-1}f_{n-1}(x) + D_n f_{n+1}(x)\right] = 0 \tag{B13a}$$

$$\begin{gathered}\left\{\frac{2n(n+\mu)}{2n+\mu+\nu} - (C_n+1)\left[\frac{1}{4}(2n+\mu+\nu+2)^2 - \frac{1}{4}(a+b-1)^2 + A_0\right] + \frac{(\nu+1)^2}{2} - \frac{(b-1)^2}{2} + A_+\right\}f_n(z) \\ -D_{n-1}\left[\frac{1}{4}(2n+\mu+\nu)^2 - \frac{1}{4}(a+b-1)^2 + A_0\right]f_{n-1}(z) - D_n\left[\frac{1}{4}(2n+\mu+\nu+2)^2 - \frac{1}{4}(a+b-1)^2 + A_0\right]f_{n+1}(z) = 0\end{gathered} \tag{B13b}$$

$$\left\{\frac{2n(n+\nu)}{2n+\mu+\nu}+(C_n-1)\left[\frac{1}{4}(2n+\mu+\nu+2)^2-\frac{1}{4}(a+b-1)^2+A_0\right]+\frac{(\mu+1)^2}{2}-\frac{(a-1)^2}{2}+A_-\right\}f_n(z)$$
$$+D_{n-1}\left[\frac{1}{4}(2n+\mu+\nu)^2-\frac{1}{4}(a+b-1)^2+A_0\right]f_{n-1}(z)+D_n\left[\frac{1}{4}(2n+\mu+\nu+2)^2-\frac{1}{4}(a+b-1)^2+A_0\right]f_{n+1}(z)=0 \quad \text{(B13c)}$$

Taking $f_0(z)=1$ these ST$^2$R$^2$ give $f_n(z)$ as a polynomial of degree $n$ in $z$, where $z\propto A_1^{-1}\left[A_0-\frac{1}{4}(a+b-1)^2\right]$ in (B13a), $z\propto A_+ +\frac{1}{2}(\nu+1)^2-\frac{1}{2}(b-1)^2$ in (B13b), and in (B13c) $z\propto A_- +\frac{1}{2}(\mu+1)^2-\frac{1}{2}(a-1)^2$. These solutions are obtained in section 3 in terms of orthogonal polynomials with continuous and discrete spectra.

It is worth noting that the ST$^2$R$^2$ (B13b) is obtained from (B13c) by the following parameter exchange symmetry

$$\mu\leftrightarrow\nu,\ a\leftrightarrow b,\ \alpha\leftrightarrow\beta,\ A_+\leftrightarrow A_-, \quad \text{(B14)}$$

in addition to $f_n\to(-1)^n f_n$, which is equivalent to $x\to -x$. Consequently, in the solution of the Jacobi-type differential equation we consider the ST$^2$R$^2$ (B13a) and either (B13b) or (B13c).